\documentclass[ 12pt]{article}

\usepackage{amssymb}
\usepackage{amsmath}
\usepackage{amssymb,amsmath}
\usepackage{float}
\usepackage{amsthm}
\newtheorem{thm}{Theorem}
\usepackage{graphicx}
\usepackage{enumerate}

\usepackage{amssymb}
\usepackage{amsmath}
\usepackage{amssymb,amsmath}
\usepackage{float}
\usepackage{amsthm}
\usepackage{multirow,array}
\usepackage{amsthm,amsmath,amssymb}
\usepackage{cases}
\usepackage{dsfont}
\usepackage[table]{xcolor}
\usepackage{color,colortbl}
\usepackage{multirow}

\usepackage{rotating}

\usepackage[margin=1.3 in]{geometry}
\usepackage{authblk}
\usepackage{amsthm,amsmath,amssymb}
\usepackage{cases}
\usepackage{dsfont}
\usepackage{multirow}
\usepackage{float}

\usepackage{graphicx}
\font\gar = eusm10 at 16pt

\def\cF{\hbox{\gar F}}

\def\sumN{{\sum_{i=1}^N}}

\def\dto{\stackrel{{d}}{\rightarrow}}
\def\one{1\!\! 1}

\def\deq{\stackrel{{d}}{\,=\,}}

\def\lawto{\stackrel{{Law}}{\longrightarrow}}
\def\disto{\stackrel{{d}}{\longrightarrow}}

\newcommand{\IR}{\mathbb R}

\newcommand{\Cs}{Cs\"org\H{o}}

\newcommand{\calF}{\mathcal F}
\newcommand{\calD}{\mathcal D}

\newtheorem{lemma}{Lemma}[section]
 \makeatletter

\@addtoreset{lemma}{thm}

\newtheorem{corollary}{Corollary}[section]
\@addtoreset{corollary}{thm}

\newtheorem{remark}{Remark}[section]
\newtheorem{prop}{Proposition}[section]
\numberwithin{equation}{section}
\begin{document}

\title{Randomized Empirical Processes and Confidence Bands via Virtual Resampling
\footnote{Research supported by a Natural Sciences and Engineering Research Council of Canada Discovery Grant of M. Cs\"{o}rg\H{o}.} }
\author{Mikl\'{o}s Cs\"{o}rg\H{o}}
\date{School of Mathematics and Statistics 1125 Colonel By Drive Ottawa, ON, Canada\\
\vspace{1 cm} In Memoriam Manny Parzen }

\maketitle

\begin{abstract}
 Let $X,X_1,X_2,\cdots$ be independent real valued random variables with a common distribution function $F$, and  consider $\{X_1,\cdots,X_N  \}$, possibly a big concrete data set, or an imaginary random sample of size $N\geq 1$ on $X$. In the latter case, or when a concrete  data set in hand is too big to be entirely processed, then the sample distribution function $F_N$ and the the population distribution function $F$ are both to be estimated. This, in this paper, is achieved via viewing $\{X_1,\cdots,X_N  \}$ as above, as a finite population of real valued random variables with $N$ labeled units, and sampling its indices $\{1,\cdots,N \}$ with replacement $m_N:= \sum_{i=1}^N w_{i}^{(N)}$ times so that for each $1\leq i \leq N$, $w_{i}^{(N)}$ is the count of number of times the index $i$ of $X_i$ is chosen in this virtual resampling process. This exposition  extends the Doob-Donsker classical theory of weak convergence of empirical processes to that of the thus created randomly weighted empirical processes when $N, m_N \rightarrow \infty$ so that $m_N=o(N^2)$.
\end{abstract}

\section{Introduction}\label{Introduction}
Let $X,X_1,X_2,\ldots$ be independent real valued random variables with a common distribution  function $F$. Consider $\{X_1,\ldots, X_N\}$, possibly a big data set of a {\sl concrete} or {\sl imaginary random sample} of size $N\ge 1$  on $X$ of a {\sl hypothetical infinite super-population}, and  define their empirical  distribution function
\begin{equation}\label{(1.1)}
F_N(x) := \sum^N_{i=1} \one(X_i \le x)/N, \quad x\in \IR,
\end{equation}
and the corresponding empirical process in this setting
\begin{eqnarray}\label{(1.2)}
\beta_N(x) &:=& \frac{1}{N^{1/2}}\sum^N_{i=1}(\one(X_i\le x)-F(x)) \nonumber\\
&=& N^{1/2}(F_N(x)-F(x)), \quad x\in  \IR,
\end{eqnarray}
where $\one(\cdot)$ is the indicator function.

\par
In case of an imaginary random sample $\{X_1,\ldots,X_N\}$ from an infinite super-population, or when a  data set is too big to be entirely processed, then the sample distribution function $F_N$ and the population distribution function $F$ are both to be estimated via taking sub-samples from an imaginary random sample, or a big data set ``in hand''.
Naturally, the same holds true for the other sample and population parameters as well, like, for example, the sample and population means and percentiles, etc. (cf.\ \Cs\ and Nasari (2015) and section \ref{CLT} in this exposition).

\par
To begin with, we view a \textit{concrete or  imaginary random sample} $\{X_1,\ldots,X_N\}$ as a  \textit{finite population} of  \textit{real  valued random variables} with $N$  \textit{labeled units}, $N\ge 1$, and sample its set of indices $\{1,\ldots,N\}$ with replacement $m_N$ times so that for each $1\le i\le N$, $w_i^{(N)}$ is the count of the number of times the index $i$ of $X_i$ is chosen in this re-sampling procedure.

\par
In view of the definition of $w_i^{(N)}$, $1\le i\le N$, in this virtual re-sampling procedure, they form a row-wise independent triangular array of random variables with $m_N := \sum^N_{i=1}w_i^{(N)}$ and, for each $N\ge 1$,

\begin{equation}\label{(1.3)}
\Big(w_1^{(N)},\ldots,w_N^{(N)}\Big) \!\deq\! \mathcal{M}ultinomial \big(\!m_N, 1/N,\ldots,1/N\!\big),
\end{equation}
i.e., the vector of the weights has a multinomial distribution of size $m_N$ with respective probabilities $1/N$.  Clearly, for each $N\ge 1$, the multinomial weights $(w_1^{(N)},\ldots, w_N^{(N)})$, by definition, are independent from the finite population of the $N$ labeled units $\{X_1,\ldots,X_N\}$.

{\bf Notations for use throughout}.  Let $(\Omega_X, \cF_x, P_X)$ denote the probability space of the i.i.d.\ random variables $X, X_1,\ldots$, and $(\Omega_w,\cF_w,P_w)$ be the probability space on which
$$(w_1^{(1)},(w_1^{(2)},w_2^{(2)}),\ldots,(w_1^{(N)},\ldots,w_N^{(N)}),\ldots),$$
are defined.  In view of the independence of these two sets of random  variables, jointly they live on the direct product probability space $(\Omega_X \times \Omega_w, \cF_X \otimes \cF_w,P_{X,w}
= P_X \times P_w)$.  For each $N\ge 1$, we also let $P_{\cdot|w}(\cdot)$ and $P_{\cdot|X}(\cdot)$ stand for the conditional probabilities given $\cF_w^{(N)} := \sigma(w_1^{(N)},\ldots,w_N^{(N)})$ and $\cF_X^{(N)} := \sigma(X_1,\ldots,X_N)$, respectively, with corresponding conditional expected values $E_{\cdot|w}(\cdot)$ and $E_{\cdot|X}(\cdot)$. Also, $E_{X,w}(\cdot)$, $E_X(\cdot)$ and $E_w(\cdot)$ will stand for corresponding  expected values in terms of $P_{X,w}(\cdot),\, P_X(\cdot)$ and $P_w(\cdot)$ respectively.

\par
We note in passing that in terms of the above notations, the distribution function $F$ of the random variable $X$ is $F(x) := P_X(X\le x)$, $x\in \IR$.

\par
Randomizing,  via the multinomial weights as in (\ref{(1.3)}), define the randomly weighted empirical process
\begin{eqnarray}\label{(1.4)}
\beta_{m_N,N}^{(1)}(x)
&:=& \frac{\sum^N_{i=1} \Bigg( \frac{w_i^{(N)}}{m_N}
                                            - \frac{1}{N}\Bigg) \one(X_i\leq x)}
{\sqrt{ \sum^N_{j=1}\Bigg(\frac{w_j^{(N)}}{m_N} - \frac{1}{N}\Bigg)^2}}\nonumber\\
&=:& \frac{F_{m_N,N}(x)-F_N(x)}
{\sqrt{ \sum^N_{j=1}\Bigg(\frac{w_j^{(N)}}{m_N} - \frac{1}{N} \Bigg)^2}} , \quad x\in \IR,
\end{eqnarray}
where
\begin{equation}\label{(1.5)}
F_{m_N,N}(x) := \sumN \frac{w_i^{(N)}}{m_N} \one(X_i\le x), \quad x\in \IR,
\end{equation}

\noindent
is the randomly weighted sample distribution function, and define as well

\begin{eqnarray}
\beta_{m_N,N}^{(2)}(x) &:=& \frac{\sum^N_{i=1} \frac{w_i^{(N)}}{m_N}\Bigg( \one(X_i \leq x)-F(x)  \Bigg) }
{\sqrt{
\sum^N_{j=1}\Bigg(\frac{w_j^{(N)}}{m_N}\Bigg)^2}},\nonumber\\
&=& \frac{F_{m_N,N}(x)-F(x)}
{\sqrt{
\sum^N_{j=1}\Bigg(\frac{w_j^{(N)}}{m_N} - \frac{1}{N}\Bigg)^2+\frac{1}{N}}}, ~~x\in\IR.
\label{(1.6)}
\end{eqnarray}
Further to these two randomly weighted empirical processes, we introduce also

\begin{eqnarray}
\beta^{(3)}_{m_N,N}(x,\theta) &:=& \frac{\sumN \Bigg( \frac{w_{i}^{(N)}}{m_N} - \frac{\theta}{N} \Bigg) \Big( \one(X_i \leq x) - F(x) \Big)} {\sqrt{\sum_{j=1}^N \Bigg( \frac{w_{i}^{(N)}}{m_N} - \frac{\theta}{N} \Bigg)^2}
 } \nonumber\\
&=& \frac{\Big(F_{m_N,N}(x)-F(x)\Big)-\theta\Big(F_N(x)-F(x) \Big) }{ \sqrt{\sum_{j=1}^N \Bigg( \frac{w_{j}^{(N)}}{m_N}- \frac{\theta}{N} \Bigg)^2 } } \nonumber\\
&=&\frac{\Big(F_{m_N,N}(x)-F(x)\Big)-\theta\Big(F_N(x)-F(x) \Big) }{ \sqrt{\sum_{j=1}^N \Bigg( \frac{w_{j}^{(N)}}{m_N}- \frac{1}{N} \Bigg)^2 + \frac{(1-\theta)^2}{N} } }, ~~ x \in \IR, \label{(1.7)}
\end{eqnarray}
where $\theta$ is a real valued constant.

\begin{remark}\label{Remark 1.1}
On letting $\theta=1$ in (\ref{(1.7)}), it reduces to (\ref{(1.4)}), while letting $\theta=0$ in (\ref{(1.7)}) yields (\ref{(1.6)}). Hence, instead of establishing the asymptotic behavior of the respective randomly weighted empirical processes of (\ref{(1.4)}) and (\ref{(1.6)}) individually on their own, it will suffice  to conclude that of $\beta^{(3)}_{m_N,N}(.,\theta)$ as in (\ref{(1.7)}) to start with.
\end{remark}

As to the weak convergence of the randomly weighted empirical process $\beta^{(3)}_{m_N,N}(.,\theta)$ as in (\ref{(1.7)}), it will be established via conditioning on the weights $\Big( w_1^{(N)}, \ldots, w_{N}^{(N)}\Big)$ as in (\ref{(1.3)}) and on assuming that $N,m_N \to \infty$ in such a way that $m_N =o(N^2)$ (cf. (\ref{(2.2)}) of Theorem \ref{THEOREM}). This, in turn, will identify the respective weak convergence of the pivotal processes $\beta^{(1)}_{m_N,N}(.)$  and $\beta^{(2)}_{m_N,N}(.)$ as that of $\beta^{(3)}_{m_N,N}(.,1)$ and $\beta^{(3)}_{m_N,N}(.,0)$ under the same conditions to a Brownian bridge $B(.)$ that, in turn, yields   Corollary \ref{Corollary 2.1}, as in Section \ref{Weak convergence}.

\par
Further to (\ref{(1.7)}), defined now the randomized empirical process $\tilde{\beta}^{(3)}_{m_N,N}(.,\theta)$ as

\begin{equation}\label{(1.8)}
\tilde{\beta}^{(3)}_{m_N,N}(x,\theta) := \sqrt{ \frac{N m_N}{N+m_N(1-\theta)^2} } \Bigg( \Big( F_{m_N,N}(x) -F(x) \Big) - \theta \Big(F_N(x) -F(x) \Big)\Bigg), ~ x\in \IR,
\end{equation}
where $\theta$ is a real valued  constant. On letting $\theta=1$ in (\ref{(1.8)}), define also $\tilde{\beta}^{(1)}_{m_N,N}(.)$ as

\begin{equation}\label{(1.9)}
\tilde{\beta}^{(1)}_{m_N,N}(x) := \tilde{\beta}^{(3)}_{m_N,N}(x,1)=\sqrt{m_N} \Bigg(F_{m_N,N}(x) - F_N (x) \Bigg), ~~ x \in \IR,
\end{equation}
and, on letting $\theta=0$, define $\tilde{\beta}^{(2)}_{m_N,N}(.)$ as

\begin{equation}\label{(1.10)}
\tilde{\beta}^{(2)}_{m_N,N}(x) := \tilde{\beta}^{(3)}_{m_N,N}(x,0)=\sqrt{\frac{N m_N}{N+m_N} } \Bigg(F_{m_N,N}(x) - F (x) \Bigg), ~~ x \in \IR.
\end{equation}

\par
Conditioning again on the weights as in (\ref{(1.3)}) and on assuming, as before, that $N, m_N \to \infty$ so that $m_N =o(N^2)$, it will be seen that the weak convergence of the virtually resampled empirical process $\tilde{\beta}^{(3)}_{m_N,N}(.,\theta)$ coincides with that of $\beta^{(3)}_{m_N,N}(.,\theta)$ (cf. (\ref{(2.3)}) of Theorem \ref{THEOREM}) and, consequently,  the respective weak convergence of the pivotal processes  $\tilde{\beta}^{(1)}_{m_N,N}(.)$ and $\tilde{\beta}^{(2)}_{m_N,N}(.)$ as in (\ref{(1.9)}) and (\ref{(1.10)})  coincide with that of $\beta^{(1)}_{m_N,N}(.)$ and $\beta^{(2)}_{m_N,N}(.)$ to a Brownian bridge $B(.)$. This, in turn, yields  Corollary \ref{Corollary 2.2} as in Section \ref{Weak convergence}.

\begin{remark}
Given the sample values of an i.i.d. sample $X_1, \cdots,X_N$
on $F$ and redrawing $m_N$ bootstrap values, define the corresponding bootstrapped empirical process \textsl{\`{a} la} $\tilde{\beta}^{(1)}_{m_N,N}$ as in (\ref{(1.9)}). In this context the latter process is studied in Section 3.6.1 of van der Vaart and Wellner (1996) (cf. $\hat{G}_{n,k}$ therein, with  $k$  bootstrap values out  of $n$ sample values of an i.i.d. sample of size $n$), where it is concluded that, in our terminology, conditioning on $X_1,\cdots, X_N$ , the ``sequence'' $\tilde{\beta}^{(1)}_{m_N,N}$ converges in distribution to a Brownian bridge in $P_X$ for every possible manner in which $m_N,N \rightarrow \infty$.

\end{remark}

\par
We note that, for $N\ge 1$,
\begin{equation}\label{(1.11)}
E_{X|w}(F_{m_N,N}(x)) = F(x),E_{w|X}(F_{m_N,N}(x))=F_N(x)
\end{equation}

and
\begin{equation}\label{(1.12)}
E_{X,w}(F_{m_N,N}(x)) = F(x), ~~\hbox{for all } x\in \IR,
\end{equation}
i.e., when conditioning on the observations $(X_1,\ldots,X_N)$ on $X$, the randomly weighted sample distribution function $F_{m_N,N}(\cdot)$ is an unbiased estimator of the sample distribution function $F_N(\cdot)$ and, when conditioning on the weights $\Big(w_1^{(N)},\ldots,w_N^{(N)}\Big)$, it is an unbiased estimator of the theoretical distribution function $F(\cdot)$.

\par
Also, for use in Section \ref{CLT} we recall that  (cf.\  \Cs\ and Nasari (2015)),  \textit{with} $N$ \textit{fixed and} $m=m_N\to\infty$,
\begin{equation}\label{(1.13)}
F_{m,N}(x) \rightarrow F_N(x)  ~~\hbox{\sl in probability}~P_{X,w}, ~~ \textit{pointwise in}~ x\in\IR,
\end{equation}

\noindent and, with $N$, $ m_N\to\infty$,
\begin{equation}\label{(1.14)}
\Big(F_{m_N,N}(x)-F_N(x)\Big)\rightarrow 0 ~~\hbox{\sl in probability}~P_{X,w}, ~~ \textit{pointwise in} ~ x\in\IR.
 \end{equation}

\par
Our approach in this exposition to the formulation of the problems in hand was inspired by
Hartley, H.O.\ and Sielken Jr., R.L.\ (1975).  A ``Super-Population Viewpoint'' for Finite Population Sampling, {\em Biometrics} {\bf 31}, 411-422. We quote  from second paragraph of 1. INTRODUCTION of this paper:

\begin{quote}
``By contrast, the super-population outlook regards the finite population of interest as a sample of size $N$ from an infinite population and regards the stochastic procedure generating the surveyor's sample of $n$ units as the following two-step procedure:

\noindent
Step 1.  Draw a ``large sample'' of size $N$  from an infinite super-population.

\noindent
Step 2. Draw a sample of size $n<N$ from the large sample of size $N$ obtained in Step 1.

\noindent
Actually, Step 1 is an imaginary step, and it is usually assumed that the resulting sample elements are independent and identically distributed.''
\end{quote}

\par
The material in this paper is organized as follows. Section \ref{Weak convergence}  spells  out weak convergence conclusions for the randomly weighted empirical processes introduced in Section \ref{Introduction}. Section \ref{Confidence bands super population} is devoted to constructing asymptotically correct size confidence bands for $F_N$ and continuous $F$, both in terms of $P_{X|w}$ and $P_{X,w}$, via virtual resampling big concrete   or imaginary random sample data sets that are viewed as samples from infinite super-populations. Section \ref{Confidence bands large sample}  and \ref{Appendix}. Appendix   deal with constructing confidence bands for continuous $F$ via virtual resampling  concrete large enough, or moderately small samples, that are to be compared to the classical Kolmogorov-Smirnov bands. Section \ref{Numerical Illustrations}   provides some numerical illustrations in view of Sections \ref{Confidence bands super population} and   \ref{Confidence bands large sample}. Section \ref{CLT} concludes  randomized central limit theorems and, consequently, confidence intervals for $F_{N}(x)$ and not necessarily continuous $F(x)$ at fixed points $x \in \mathds{R}$ via virtual resampling. All the proofs are given in Section \ref{Proofs}.

\section{Weak convergence of randomly weighted empirical processes}\label{Weak convergence}
In view of $\{X_1,\ldots,X_N\}$ possibly being a  \textit{big data set}, or an  \textit{imaginary random sample of size} $N\ge 1$ on $X$ of a  \textit{hypothetical infinite
super-population}, the weak convergence of the randomized empirical processes
$\{\beta^{(i)}_{m_N,N}(x), \beta^{(3)}_{m_N,N}(x, \theta)\ , x\in \IR\}$, $i=1,2,$, $\theta \in \IR$, $\{\tilde{\beta}^{(i)}_{m_N,N}(x), \tilde{\beta}^{(3)}_{m_N,N}(x, \theta)\ , x\in \IR\}$, $i=1,2,$, $\theta \in \IR$,   is  studied mainly  for the sake of forming  \textit{asymptotically exact size confidence bands} for $F_N$ and $F$, based on $\{\beta^{(i)}_{m_N,N}(x), x \in \IR \}$, $i=1,2$,  and  $\{\tilde{\beta}^{(i)}_{m_N,N}(x), x \in \IR \}$, $i=1,2$, with sub-samples of size $m_N<N$, (cf. Section \ref{Confidence bands super population}).

\par
For the right continuously defined distribution function $F(x):= P_X(X\le x)$, we define its left continuous inverse (quantile function) by
\begin{equation}\nonumber
F^{-1}(t) := \inf  \left\{x\in \IR : F(x) \ge t\right\},
\qquad  0<t\le 1, ~F^{-1}(0)=F^{-1}(0+).
\end{equation}
Thus, in case of a continuous distribution function $F$, we have
$$
F^{-1}(t) =\inf\left\{x\in\IR:F(x)=t\right\}\hbox{~and~} F(F^{-1}(t))=t\in[0,1].
$$
Consequently, the  random variable  \ $F(X)$ is uniformly distributed on the unit interval [0,1]:

\begin{eqnarray*}\nonumber
P_X(F(X)\ge t)=P_X(X \ge F^{-1}(t)) &=& 1-P_X(X\le F^{-1}(t)) \\
&=& 1-F(F^{-1}(t))=1-t,
 \ 0\le t\le 1.
\end{eqnarray*}

Hence, if $F$ is continuous, then the classical empirical process (cf.(\ref{(1.2)})) in this setting becomes
\begin{eqnarray}
\beta_N(F^{-1}(t))
&=& \frac{1}{N^{1/2}} \sum^N_{i=1}\Bigg(\one(X_i \le F^{-1}(t))-F(F^{-1}(t))\Bigg)\nonumber \\
&=& \frac{1}{N^{1/2}} \sum^N_{i=1}\Bigg(\one(F(X_i) \le  t)-t\Bigg), ~0\le t\le 1,
\label{(2.1)}
\end{eqnarray}
the uniform empirical process of the independent uniformly distributed random variables  $F(X_1),\ldots,F(X_N)$, $N\geq 1$.

\par
Accordingly, when $F$ is continuous, the weak convergence of $\beta_N(x)$, $x\in \IR$, as in (\ref{(1.2)}) can be established via that of $\beta_N(F^{-1}(t))$, $0\le t\le 1$, as in (\ref{(2.1)}),  and, as $N\to\infty$, we have (Doob (1949), Donsker (1952))

\begin{equation}\label{new (2.1)}
\beta_N(F^{-1}(\cdot)) \lawto B(\cdot) \hbox{~~on~~} (D,\calD, ~\Vert~\Vert),
\end{equation}
with notations as in our forthcoming Theorem \ref{THEOREM},  where $B(\cdot)$ is a Brownian bridge on $[0,1]$, a Gaussian process with covariance function  $E B(s)B(t)=s \wedge t-st$, that in terms of a standard Brownian motion $\{W(t),\, 0\le t<\infty\}$ can be defined as
\begin{equation}\nonumber
\{B(t),\, 0\le t\le 1\} = \{W(t)-tW(1), \, 0\le t\le 1\}.
\end{equation}

\par
Mutatis mutandis, when $F$ is continuous, similar conclusions hold true for obtaining the weak convergence of the randomly weighted empirical processes  as in (\ref{(1.4)}), (\ref{(1.6)}), (\ref{(1.7)}),  (\ref{(1.8)}), (\ref{(1.9)}) and (\ref{(1.10)}), respectively, via establishing that of their corresponding uniform  versions that, for the sake of theorem proving, are defined in terms of the  independent uniformly distributed random variables  $F(X_1),\ldots,F(X_N)$, $N\ge 1$.

\begin{thm}\label{THEOREM}
Let $X,X_1,\ldots$ be real valued i.i.d.\ random variables on $(\Omega_X, \calF_X, P_X)$.  Assume that $F(x)\!=$ $P_X(X\!\le\! x)$, $x\in\IR$, is a continuous distribution  function.  Then, relative to the conditional distribution $P_{X|w}$, if $N,m_N\to\infty$ ~so that~ $m_N=o(N^2)$, via (\ref{(1.7)}), with $\theta \in \IR$, we have

\begin{equation}\label{(2.2)}
\beta^{(3)}_{m_N,N}\bigg(F^{-1}(\cdot),\theta\bigg) \lawto B(\cdot) \hbox{ on }(D,\calD,\Vert~\Vert),~~~~ \textrm{in~ probability} ~ P_w,
\end{equation}
and, via (\ref{(1.8)}), with $\theta \in \IR$,  we get
\begin{equation}\label{(2.3)}
\tilde{\beta}^{(3)}_{m_N,N}\bigg(F^{-1}(\cdot), \theta\bigg) \lawto B(\cdot) \hbox{ on }(D,\calD,\Vert~\Vert), ~ ~ \textrm{in ~ probability}~P_w,
\end{equation}
where $B(\cdot)$ is a Brownian bridge on $[0,1]$, $\calD$ denotes the $\sigma$-field generated by the finite dimensional subsets of $D=D[0,1]$, and $\Vert~\Vert$ stands for the uniform metric for real valued functions on $[0,1]$, i.e., in both cases we have weak convergence on $(D,\calD,\Vert~\Vert)$ in probability $P_w$
%via convergence in distribution of all finite dimensional distributions and tightness
in terms of $P_{X|w}$.
\end{thm}

\begin{remark}
We note in passing that, suitably stated, the conclusions of Theorem \ref{THEOREM} continue to hold true in terms of an arbitrary distribution function $F(.)$ as well. Namely, in the latter case, let $F(.)$ be defined to be right continuous, and define its left continuous inverse $F^{-1}(.)$ as before. Then, relative to the conditional distribution $P_{X|w}$, if $N,m_N \to \infty$ so that $m_N=o(N^2)$, in lieu of (\ref{(2.2)}), in probability $P_w$ we have

\begin{equation}\nonumber
\beta^{(3)}_{m_N,N}\bigg(F^{-1}(\cdot), \theta\bigg) \lawto B\big(F\big(F^{-1}(\cdot)) \big)\hbox{ on }(D,\calD,\Vert~\Vert),
\end{equation}
and,  in lieu of (\ref{(2.3)}), in probability $P_w$ we have
\begin{equation}\nonumber
\tilde{\beta}^{(3)}_{m_N,N}\bigg(F^{-1}(\cdot), \theta\bigg) \lawto B\big(F\big(F^{-1}(\cdot))\big) \hbox{ on }(D,\calD,\Vert~\Vert).
\end{equation}
Consequently, limiting functional laws will depend on $F$, unless it is continuous.

\end{remark}

\begin{corollary}\label{Corollary 2.1}
As  $N,m_N \to\infty$ so that $m_N = o(N^2)$,    via (\ref{(2.2)}) with $\theta \in \IR$ we conclude in probability $P_w$

\begin{equation}\label{(2.4-1)}
P_{X|w}\big(h(\beta^{(3)}_{m_N,N}(F^{-1}(\cdot),\theta)\le y) \big)\to P(h(B(\cdot))\le y)=:G_{h(B(.))} (y), ~~ y\in \IR,
\end{equation}
on letting $\theta=1$ in (\ref{(2.4-1)}) and recalling Remark \ref{Remark 1.1},

\begin{equation}\label{(2.4)}
P_{X|w}\big( h(\beta^{(1)}_{m_N,N}(F^{-1}(\cdot))\le y) \big) \to
 P(h(B(\cdot))\le y) =:
 G_{h(B(\cdot))}(y), ~~y\in \IR,
\end{equation}
and, on letting $\theta=0$ in (\ref{(2.4-1)}) and recalling Remark \ref{Remark 1.1},

\begin{equation}\label{(2.5)}
P_{X|w} \big(h(\beta^{(2)}_{m_N,N}(F^{-1}(\cdot))\le y) \big) \to P(h(B(\cdot))\le y)
=: G_{h(B(\cdot))}(y), ~~y\in \IR,
\end{equation}
at all points of continuity of the  distribution function $G_{h(B(\cdot))}(\cdot)$ for all functionals  $h:D\to\IR$ that are $(D,\calD)$ measurable and $\Vert ~\Vert$-continuous, or  $\Vert ~\Vert$-continuous, except at points forming a set of  measure zero on $(D,\calD)$ with respect to the measure  generated by $\{B(t),\, 0\le t\le 1\}$.
\end{corollary}

\par
Consequently, by a bounded convergence theorem as spelled out in Lemma 1.2 of S. Cs\"{o}rg\H{o} and Rosalsky (2003),  as $N,m_N\to \infty$ so that $m_N=o(N^2)$, respectively via (\ref{(2.4-1)}) and   (\ref{(2.4)})  we conclude  also

\begin{equation}\label{(2.6-1)}
P_{X,w}\big( h(\beta^{(3)}_{m_N,N}(F^{-1}(\cdot), \theta))\le y\big) \to
G_{h(B(\cdot))}(y), ~~y\in \IR,
\end{equation}
and
\begin{equation}\label{(2.6)}
P_{X,w}\big(h(\beta^{(1)}_{m_N,N}(F^{-1}(\cdot)))\le y \big) \to
G_{h(B(\cdot))}(y), ~~y\in \IR,
\end{equation}
a remarkable extension of (\ref{(1.14)}), \verb""
and,  via (\ref{(2.5)}) we arrive at having as well

\begin{equation}\label{(2.7)}
P_{X,w}\big(h(\beta^{(2)}_{m_N,N}(F^{-1}(\cdot))) \le y \big) \to
G_{h(B(\cdot))}(y), ~~y\in \IR,
\end{equation}
at all points of continuity of the distribution function $G_{h(B(\cdot))}(\cdot)$ for all functionals $h:D\to\IR$ as spelled out right after (\ref{(2.5)}) above.

\par
In the sequel, we will make use of taking $h$ to be the sup functional on $D[0,1]$.

\par
In view of (\ref{(2.4)}) and (\ref{(2.6)}), {\sl as $N,m_N\to\infty$ so that}  $m_N=o(N^2)$, with $\dto$ standing for convergence in distribution, we have
\begin{eqnarray}
\sup_{x\in\IR} \left|\beta^{(1)}_{m_N,N}(x)\right|
 &=& \frac{1}{\sqrt{\sum_{j=1}^N (\frac{w_j^{(n)}}{m_N}-\frac{1}{N} )^2}} \sup_{x \in \IR} \left| F_{m_N,N}(x)-F_{N}(x) \right| \nonumber\\
&=&\sup_{0\le t\le 1}\left|\beta^{(1)}_{m_N,N}(F^{(-1)}(t))\right|
\dto~ \sup_{0\le t\le 1} |B(t)|\label{(2.8)}
\end{eqnarray}
{\sl both in terms of $P_{X|w}$ and} $P_{X,w}$, while in view of (\ref{(2.5)}) and (\ref{(2.7)}), {\sl  we conclude
\begin{eqnarray}
\sup_{x\in\IR} \left|\beta^{(2)}_{m_N,N}(x)\right|
 &=& \frac{1}{\sqrt{\sum_{j=1}^N (\frac{w_j^{(n)}}{m_N}-\frac{1}{N} )^2+\frac{1}{N}}} \sup_{x \in \IR} \left| F_{m_N,N}(x)-F(x) \right| \nonumber\\
&=& \sup_{0\le t\le 1} \left|\beta^{(2)}_{m_N,N}(F^{-1}(t)))\right|
\dto ~\sup_{0\le t\le 1} |B(t)| \label{(2.9)}
\end{eqnarray}
both in terms of $P_{X|w}$ and $P_{X,w}$.}%end of sl

\par
As a consequence of (\ref{(2.4-1)}) and (\ref{(2.6-1)}), as $N,m_n \to \infty$ so that $m_N=o(N^2)$, for $\beta_{m_N,N}^{(3)}(x,\theta)$ as in (\ref{(1.7)}) with $F(.)$ continues, we conclude

\begin{eqnarray}
\sup_{x \in \IR} \Big| \beta_{m_N,N}^{(3)} \big(x,\theta  \big) \Big| &=& \frac{  \sup_{x \in \IR }  \Big| \Big( F_{m_N,N}(x)-\theta F_N(x)\Big) - (1-\theta)F(x)    \Big|  }{ \sqrt{\sum_{j=1}^N  \big( \frac{w_{j}^{(N)}}{m_N}-\frac{1}{N}  \big)^2 +\frac{(1-\theta)^2}{N}  }   }\nonumber\\
&=& \sup_{0 \leq t \leq 1} \Big| \beta_{m_N,N}^{(3)} \big(F^{-1}(t),\theta  \big)  \Big| \to \sup_{0 \leq t \leq 1} \big| B(t) \big| \label{(2.10-1)}
\end{eqnarray}
both in terms of $P_{X|w}$ and $P_{X,w}$.

\begin{remark} \label{Remark 2.2-1}
We note in passing that on letting $\theta=1$, the statement of (\ref{(2.10-1)})  reduces  to that of (\ref{(2.8)}), while on letting $\theta=0$, we arrive at that of (\ref{(2.9)}). With $\{ \theta \in \IR| \theta \neq 1 \}$ in (\ref{(2.10-1)}), we have $F_{m_N,N}(x) -\theta F_N (x)$ estimating $(1-\theta) F(x)$ uniformly in $x \in \IR$, a companion to $F_{m_N,N}(x)$ also estimating $F(x)$ uniformly in $x \in \IR$ as in (\ref{(2.9)}), as well as to $F_N (x)$ alone estimating $F(x)$ uniformly in $x \in \IR$ as  right below  in (\ref{(2.10)}).
\end{remark}

\par
Naturally, for the classical empirical process $\beta_N(\cdot)$ (cf.\ (\ref{(1.2)}) and (\ref{(2.1)}) with $F(x)=P_X(X\le x)$ continuous), as $N\to\infty$, via (\ref{new (2.1)}) we arrive at
\begin{equation}\label{(2.10)}
\sup_{x\in\IR} |\beta_N(x)| = \sup_{0\le t\le 1} |\beta_N(F^{-1}(t))| \dto \sup_{0\le t\le 1} |B(t)|
\end{equation}
in terms of $P_X$.

\begin{corollary}\label{Corollary 2.2}
As $N,m_N \to \infty$ so that $m_N=o(N^2)$,  via (\ref{(2.3)}) and (\ref{(1.8)}) with $\theta \in \IR$, we conclude in probability $P_w$

\begin{equation}\label{(2.12-1)}
P_{X|w} \Big(  h\Big(  \tilde{\beta}_{m_N,N}^{(3)} \big( F^{-1}(\cdot), \theta \big) \Big) \leq y \Big) \to P\Big(  h\Big(  B(.) \Big)\leq y \Big)
= G_{h\big( B(\cdot) \big)}(y),~ y \in \IR,
\end{equation}
on letting $\theta=1$  in (\ref{(1.8)}) and recalling (\ref{(1.9)})

\begin{equation}\label{(2.12)}
P_{X|w} \Big(  h\Big(  \tilde{\beta}_{m_N,N}^{(1)} \big( F^{-1}(\cdot) \big) \Big) \leq y \Big) \to P\Big(  h\Big(  B(.) \Big)\leq y\Big)
= G_{h\big( B(\cdot) \big)}(y),~ y \in \IR,
\end{equation}
and on letting $\theta=0$, in (\ref{(1.8)}) and recalling (\ref{(1.10)})

\begin{equation}\label{(2.13)}
P_{X|w} \Big(  h\Big(  \tilde{\beta}_{m_N,N}^{(2)} \big( F^{-1}(\cdot) \big) \Big) \leq y \Big) \to P\Big(  h\Big(  B(.) \Big)\leq y\Big)= G_{h\big( B(\cdot) \big)}(y),~ y \in \IR,
\end{equation}
at all points of continuity of the distribution function $G_{h\big( B(\cdot) \big)}(\cdot)$ for all functionals $h:D\to \IR$ that are $(D,\mathcal{D})$ measurable and $\|~ \|$-continuous, or $\|~ \|$-continuous, except at points forming a set of measure zero on $(D,\mathcal{D})$ with respect to the measure generated by $\{B(t),~ 0 \leq t \leq 1 \}$.
\end{corollary}

\par
Consequently, again in view of Lemma 1.2 of S. Cs\"{o}rg\H{o} and Rosalsky (2003),  as $N,m_N \to \infty$ so that $m_N=o(N^2)$, respectively  via   (\ref{(2.12-1)}),    (\ref{(2.12)}) and (\ref{(2.13)}),  we conclude also

\begin{equation}\label{(2.14-1)}
P_{X,w}\Bigg( h\Big( \tilde{\beta}^{(3)}_{m_N,N}\big( F^{-1}(\cdot),\theta\big) \Big) \leq y   \Bigg) \to G_{h\big(B(\cdot) \big)}(y), ~ y \in \IR,
\end{equation}

\begin{equation}\label{(2.14)}
P_{X,w}\Bigg( h\Big( \tilde{\beta}^{(1)}_{m_N,N}\big( F^{-1}(\cdot)\big) \Big) \leq y   \Bigg) \to G_{h\big(B(\cdot) \big)}(y), ~~ y \in \IR,
\end{equation}
and

\begin{equation}\label{(2.15)}
P_{X,w}\Bigg( h\Big( \tilde{\beta}^{(2)}_{m_N,N}\big( F^{-1}(\cdot)\big) \Big) \leq y   \Bigg) \to G_{h\big(B(\cdot) \big)}(y), ~~ y \in \IR,
\end{equation}
at all points of continuity of the distribution function $G_{h\big(B(\cdot) \big)}(\cdot)$ for all functionals $h:D\to \IR$ as spelled out right after (\ref{(2.13)}) above.

\par
On taking $h$ to be the sup functional on $D[0,1]$ in (\ref{(2.12-1)}) - (\ref{(2.15)}), as $N,m_N \to \infty$ so that $m_N=o(N^2)$, we conclude, both in terms of $P_{X|w}$ and $P_{X,w}$,

\begin{eqnarray}
\sup_{x \in \IR} \Big| \tilde{\beta}^{(3)}_{m_N,N}(x,\theta) \Big| &=& \sqrt{\frac{N m_N}{N+m_N(1-\theta)^2} }  \sup_{x \in \IR} \Big|\Big(F_{m_N,N}(x)-\theta F_{N}(x)\Big) -(1-\theta) F(x) \Big|\nonumber\\
&=& \sup_{0 \leq t \leq 1} \Big| \tilde{\beta}^{(3)}_{m_N,N}\big( F^{-1}(t),\theta \big) \Big| \dto \sup_{0\leq t \leq 1} \Big|B(t) \Big|, \label{(2.16-1)}\\
\sup_{x \in \IR} \Big| \tilde{\beta}^{(1)}_{m_N,N}(x) \Big| &=& \sqrt{m_N} \sup_{x \in \IR} \Big|F_{m_N,N}(x)-F_{N}(x) \Big|\nonumber\\
&=& \sup_{0 \leq t \leq 1} \Big| \tilde{\beta}^{(1)}_{m_N,N}\big( F^{-1}(t) \big) \Big| \dto \sup_{0\leq t \leq 1} \Big|B(t) \Big|, \label{(2.16)}\\
\sup_{x \in \IR} \Big| \tilde{\beta}^{(2)}_{m_N,N}(x) \Big| &=& \sqrt{\frac{N m_N}{N+m_N}} \sup_{x \in \IR} \Big|F_{m_N,N}(x)-F(x) \Big|\nonumber\\
&=& \sup_{0 \leq t \leq 1} \Big| \tilde{\beta}^{(2)}_{m_N,N}\big( F^{-1}(t) \big) \Big| \dto \sup_{0\leq t \leq 1} \Big|B(t) \Big|. \label{(2.17)}
\end{eqnarray}

\begin{remark}\label{Remark 2.3}
Mutatis mutandis, the conclusions of Remark \ref{Remark 2.2-1} continue to be valid concerning (\ref{(2.16-1)}) - (\ref{(2.17)}) and (\ref{(2.10)}).
\end{remark}

\section{Confidence bands for empirical and theoretical distributions via virtual resampling big concrete or imaginary random sample data sets that are  viewed as samples from  infinite super-populations}
\label{Confidence bands super population}
Let $\{X_1,\dots,X_N \}$ be a big i.i.d. data set on $X$, or an imaginary random sample (a finite population) of size $N\geq 1$ on $X$, of a hypothetical infinite super-population with distribution function $F(x)=P_X(X \leq x)$, $x \in \IR$. As in Section \ref{Weak convergence}, we assume throughout this section that $F(.)$ is a continuous distribution function.

\par
Big data sets in this section refer to having too many data in one random sample as above that in some cases need to be stored on several machines, on occasions even on thousands of machines. In some cases processing  samples of this size may be virtually impossible or, simply, even undesirable to make use of the whole sample.

\par
To deal with this problem, as well as with that of sampling from a finite population, we explore and exploit  the virtual resampling method of taking sub-samples from the original big data set, or finite population, of size $N$ so that only the  reduced number of the picked elements $m_N$ of the original sample  are to be used to infer about the parameters of interest of the big data set, or  of a finite population, as well as those  of their super-population. This can be done by generating a realization of multinomial  random variables $\big( w_{1}^{(N)}, \ldots, w_{N}^{(N)} \big)$ of size $m_N=\sum_{i=1}^N w_{i}^{(N)}$, independently from the data (cf. (\ref{(1.3)})) so that $m_N << N$. The generated weights are to be put in a one-to-one correspondence with the indices of the members of the original concrete or imaginary  data set. Then, \textit{only} those data in the set in hand are to be observed whose corresponding weights are not zero.

\par
We note in passing that in  the case when the sub-sample size $m_N$ and the sample size $N$ are so that $m_N/N\leq 0.05$, then for each $i$, $1\leq i \leq N$, $w_{i}^{(N)}$ is either zero or one almost all the time. Thus, in this case, our virtual resampling with replacement method as in (\ref{(1.3)}), practically reduces to reduction via virtual resampling \textit{without} replacement.

\par
In the present context of viewing big data sets and finite populations as random samples and imaginary random samples respectively from an infinite super-population, we are to construct asymptotically exact size confidence bands for both $F_N (.)$ and $F(.)$, via our randomized empirical processes. To achieve this goal, we may use the respective conclusions of (\ref{(2.8)}) and (\ref{(2.9)}), or, asymptotically equivalently (cf. Lemma \ref{Lemma 6.2}), those of  (\ref{(2.16)}) and (\ref{(2.17)}). In view of the more familiar  appearance  of the respective  norming sequences in the   conclusions of  (\ref{(2.16)}) and (\ref{(2.17)}), in this section we are to make use of the latter two that are also more convenient for doing calculations.

Consider  $\tilde{\beta}^{(1)}_{m_N,N}(x)$, as in (\ref{(1.9)}) and define the event, a confidence set for $F_{N}(.)$,

\begin{equation}\label{(3.1)}
\mathcal{A}^{(1)}_{m_N,N}(c_{\alpha}):=\big\{ L^{(1)}_{m_N,N}(x)\! \le\! F_N(x)\! \le\! U^{(1)}_{m_N,N}(x),\forall x\!\in\!\IR\big\},
\end{equation}
where~
\begin{eqnarray}
L^{(1)}_{m_N,N} (x) &:=& F_{m_N,N}(x)-c_\alpha/\sqrt{m_N}, \label{(3.2)}\\
U^{(1)}_{m_N,N}(x) &:=& F_{m_N,N}(x)+c_\alpha/\sqrt{m_N}, \label{(3.3)}
\end{eqnarray}
and, given $\alpha\in (0,1)$, $c_\alpha$ is the $(1-\alpha)$th quantile of the distribution function of the random variable \ $~\sup_{0\le t\le 1}|B(t)|$, i.e.,

\begin{equation}\label{(3.4)}
K(c_\alpha) := P\Bigg(\sup_{0\le t\le 1}|B(t)|\le c_\alpha\Bigg)= 1-\alpha.
\end{equation}

\par
Then, as $N,m_N \to \infty$ so that $m_N=o(N^2)$,  by (\ref{(2.16)})  we obtain an asymptotically correct $(1-\alpha)$ size confidence set for $F_N (.)$, both in terms of $P_{X|w}$ and $P_{X,w}$, that via $P_{X,w}$ reads as follows

\begin{equation*}
P_{X,w}(\mathcal{A}^{(1)}_{m_N,N}(c_{\alpha}))=P_{X,w} \Big( \sup_{x \in \IR} \big| \tilde{\beta}_{m_N,N}^{(1)}(x) \big|\leq c_{\alpha}\Big) \longrightarrow 1-\alpha.
\end{equation*}

\par
Consequently, since $F_{N}(x)$, $x \in \IR$, takes on its values in $[0,1]$, it follows that, under the same conditions,

\begin{equation}\label{(3.5)}
\Big\{ \max\Big( 0, L_{m_N,N}^{(1)}(x)   \Big) \leq F_N (x) \leq \min\Big( 1, U_{m_N,N}^{(1)}(x) \Big), ~ \forall x \in \mathds{R} \Big\}
\end{equation}
is an asymptotically correct $(1-\alpha)$ size confidence band for $F_{N}(\cdot)$, both in terms of $P_{X|w}$ and $P_{X,w}$.

\par
To illustrate the reduction of the number of data that is needed for covering $F_N(x)$ for all $x\in\IR$ in case of a big data set or a finite population, say of size $N=10^4$, on taking $m_N=\sqrt N$, we have

\begin{equation}\nonumber
m_{10^4} = \sum^{10^4}_{i=1} w_i^{  (10^4)} = (10^4)^{1/2} = 100,
\end{equation}
where the random  multinomially distributed weights  $\Big(w_1^{(10^4)},\ldots,w_{10^4}^{(10^4)}\Big)$ are generated independently from the data $\{X_1,\ldots,X_{10^4}\}$ with respective probabilities~ $1/10^4$,~ i.e.,

\begin{equation}\nonumber
\Big(w_1^{(10^4)}, \ldots,w_{10^4}^{(10^4)}\Big) \deq    \mathcal{M}ultinomial  \Big(100; 1/{10^4},\ldots,1/{10^4}\Big).
\end{equation}
These multinomial weights, in turn, are used to construct an asymptotically correct $(1-\alpha)$ size confidence set {\sl \`a la} (\ref{(3.1)}) and (\ref{(3.5)}), covering the unobserved sample distribution function $F_N(x)$ uniformly in $x\in\IR$.

\par
Consider now $\tilde{\beta}^{(2)}_{m_N,N}$, as in (\ref{(1.10)}), and define the event, a confidence set for $F(.)$,

\begin{equation}\label{(3.6)}
\mathcal{A}^{(2)}_{m_N,N}(c_{\alpha}):=\big\{ L^{(2)}_{m_N,N}(x)\! \le\! F(x)\! \le\! U^{(2)}_{m_N,N}(x),\forall x\!\in\!\IR\big\},
\end{equation}
where~
\begin{eqnarray}
L^{(2)}_{m_N,N} (x) &:=& F_{m_N,N}(x)-c_\alpha \sqrt{1/m_N + 1/N} \nonumber\\
&=& F_{m_N,N}(x)-c_\alpha \sqrt{\frac{1+m_N/N}{m_N}}, \label{(3.7)}\\
U^{(2)}_{m_N,N}(x) &:=& F_{m_N,N}(x)+c_\alpha\sqrt{1/m_N + 1/N}\nonumber\\
&=& F_{m_N,N}(x)+c_\alpha \sqrt{\frac{1+m_N/N}{m_N}},  \label{(3.8)}
\end{eqnarray}
and, given $\alpha\in (0,1)$, $c_\alpha$ is again as in  (\ref{(3.4)}).

\par
Thus, as $N,m_N \to \infty$ so that $m_N=o(N^2)$, by (\ref{(2.17)}), we now obtain an asymptotically correct $(1-\alpha)$ size confidence set  for $F(.)$, both in terms of $P_{X|w}$ and $P_{X,w}$, that via $P_{X,w}$ reads as follows

\begin{equation}%\label{(3.9)}
P_{X,w}\big(\mathcal{A}^{(2)}_{m_N,N}(c_{\alpha})\big) = P_{X,w} \big(  \sup_{x \in \IR} \big|\tilde{\beta}^{(2)}_{m_N,N}(x) \big| \leq c_{\alpha} \big)\longrightarrow 1-\alpha.
\end{equation}

\par
Consequently, since $F(x)$, $x \in \IR$, takes on its values in $[0,1]$, it follows that, under the same conditions,

\begin{equation}\label{(3.9)}
\big\{ \max\big( 0,L^{(2)}_{m_N,N} (x) \big)\leq  F(x) \leq  \min\big( 0,U^{(2)}_{m_N,N} (x) \big), ~ \forall x \in \IR \big\}
\end{equation}
is an asymptotically correct $(1-\alpha)$ size confidence band for $F(.)$, both in terms of $P_{X|w}$ and $P_{X,w}$.

\par
As to the sub-samples of size $m_N$, $N\geq 1$, that are to be used for covering $F_N(.)$ and $F(.)$, for all $x \in \IR$, respectively  as in (\ref{(3.5)}) and (\ref{(3.9)}), we have $N,m_N \to \infty$ so that $m_N=o(N^2)$ in both cases. Thus we may for example consider having $N,m_N \to \infty$ so that $m_N=o(N)$, say $m_N=N^{\varepsilon}$ with $0< \varepsilon <1$. Then the respective lower and upper bounds in (\ref{(3.1)}) and (\ref{(3.6)}) for covering $F_N(.)$ and $F(.)$ respectively via (\ref{(3.5)}) and (\ref{(3.9)}) will eventually coincide, and the asymptotically correct $(1-\alpha)$ confidence bands therein   will practically be of equal width.  For instance, using the above illustrative example right after (\ref{(3.5)}), where $N=10^4$ and $m_N=m_{10^4}=100$, in the event $\mathcal{A}_{m_N,N}^{(1)}(c_{\alpha})$, as in (\ref{(3.1)}), in this case $c_{\alpha}$ is multiplied by $\sqrt{1/100}$, while in the event $\mathcal{A}_{m_N,N}^{(2)}(c_{\alpha})$ as in (\ref{(3.6)}), $c_{\alpha}$ is multiplied by $\sqrt{1/100 + 1/10^4}$.

\par
In view of having $m_N,N \to \infty$ so that $m_N=o(N^2)$ in both (\ref{(3.5)}) and (\ref{(3.9)}), we may also consider the case $m_N=O(N)$ as $N \to \infty$, and can thus, e.g., also have $m_N=c N$ with a small constant $0< c <<1$, in the context of this section.

\begin{remark}\label{Remark 3.1}
We can also make use of the functional $\sup_{x \in \IR}  \big| \tilde{\beta}^{(2)}_{m_N,N} (x) \big|$
as in (\ref{(2.17)}) for goodness of fit tests for $F$ against general alternatives in our present context of virtual resampling big data sets and finite populations when  they are viewed as samples from infinite super-populations with continuous $F$. Namely, for testing the null hypothesis $H_0:~ F=F_0$, where $F_0$ is a given continuous distribution function, we let $F=F_0$ in (\ref{(2.17)}), and reject $H_0$ in favor of the alternative $H_1:~  F\neq F_0$, for large values of the thus obtained statistic at significance level $\alpha \in (0,1)$ as $N,m_N \to \infty$ so that $m_N=o(N^2)$. Thus, in view of (\ref{(2.17)}), as $N,m_N \to \infty$ so that $m_N=o(N^2)$, an  asymptotic size $\alpha \in(0,1)$ Kolmogorov type test for $H_0$ versus $H_1$ has the rejection region, both in terms of $P_{X|w}$ and $P_{X,w}$,

\begin{equation}\label{(3.10)}
\sqrt{\frac{N m_N}{N+m_N}} \sup_{x \in \IR} \big|F_{m_N,N}(x) - F_{0}(x)  \big| \geq c_{\alpha},
\end{equation}
where $c_{\alpha}$ is as in (\ref{(3.4)}).
\end{remark}

\begin{remark}\label{Remark 3.2}
In view of the conclusions of (\ref{(2.13)}) and (\ref{(2.15)}), that are asymptotically equivalent to those of (\ref{(2.4)}) and (\ref{(2.7)}), we may also consider other functionals $h(.)$ for goodness of fit tests for $F$ against general alternatives in our present context of virtual resampling big data sets and finite populations as right above in Remark \ref{Remark 3.1}. For example, based  on (\ref{(2.13)}) and (\ref{(2.15)}), as $N,m_N \to \infty$ so that $m_N=o(N^2)$, we have, both in terms of $P_{X|w}$ and $P_{X,w}$ ,

\begin{eqnarray}\label{(3.11)}
&&\frac{N m_N}{N+m_N}  \int^{+\infty}_{-\infty}   \Big( F_{m_N,N} (x) - F(x)  \Big)^2 dF(x)
=  \int^{1}_{0} \Big( \tilde{\beta}_{m_N,N}^{(2)} \big( F^{-1}(t) \big)  \Big)^2 dt \dto  \int^{1}_{0} B^2(t) dt.\nonumber\\
&&
\end{eqnarray}
Thus, in view of (\ref{(3.11)}), as $N,m_N \to \infty$ so that $m_N=o(N^2)$, an asymptotic size $\alpha \in (0,1)$ Cram\'{e}r-von Mises-Smirnov type test for testing $H_0$ versus $H_1$ as in Remark \ref{Remark 3.1} has the rejection region

\begin{equation}\label{(3.12)}
\omega_{N,m_N}^2:= \frac{N m_N}{N+m_N} \int^{+\infty}_{-\infty}   \Big( F_{m_N,N} (x) - F_{0}(x)  \Big)^2 dF_{0}(x) \geq \nu_{\alpha}
\end{equation}
where $\nu_{\alpha}$ is the  $(1-\alpha)$th  quantile of the distribution function of the random variable $\omega^2=\int^{1}_{0} B^2(t) dt$, i.e.,

\begin{equation}\label{(3.13)}
V(\nu_{\alpha}):= P(\omega^2 \leq \nu_{\alpha})=1-\alpha.
\end{equation}

\end{remark}

\section{Confidence bands for  theoretical distributions   via   virtual resampling from  large enough, or moderately small,     samples}\label{Confidence bands large sample}
When all $N$ observables of large enough, or moderately small, samples are available to be processed, then
(\ref{(2.10)}) yields  the  asymptotically exact $(1-\alpha)$ size classical Kolmogorov confidence band for a continuous distribution function $F(x)$. Namely, as $N \to \infty$, with  probability $(1-\alpha)$, $\alpha \in (0,1)$, we have

\begin{equation}\label{(4.1)}
\big\{  \max\big( 0,F_N (x) - c_{\alpha}/\sqrt{N} \big) \leq F(x) \leq \min\big(1, F_{N}(x) +c_{\alpha}/\sqrt{N} \big), \ \forall x \in \IR   \Big\}
\to 1-\alpha,
\end{equation}
where $c_{\alpha}$ is as in (\ref{(3.4)}).

\par
Next, in the present context, in the asymptotically correct $(1-\alpha)$ size confidence band for $F(.)$  as in (\ref{(3.9)}) that obtains  as $N,m_N \rightarrow \infty$ so that $m_N=o(N^2)$, and is valid both in terms of $P_{X|w}$ and $P_{X,w}$, we may for example let $m_N=N$ in case of having large enough samples, and $m_N=N^{\varepsilon}$ with $1<\varepsilon<2$, when having moderately small samples, and compare the thus obtained bands in (\ref{(3.9)}) to that of the classical one in (\ref{(4.1)}).

\par
Also, the conclusion of (\ref{(3.5)}) provides an asymptotically correct $(1-\alpha)$ size confidence band for the observable $F_{N}(.)$ itself in a similar way, via the also observable $F_{m_{N},N}(.)$, say as above, with $m_N=N$ in case of moderately large samples and, with $m_N=N^{\varepsilon}$, $1< \varepsilon <2$, when having only moderately small samples.

\par
The goodness of fit tests as in (\ref{(3.10)})  and (\ref{(3.12)}) can also be used for example with $m_N$ as above, when all observables of large enough, or moderately small, samples are available to be processed.

\par
Furthermore, as noted in Remarks \ref{Remark 2.2-1} and  and \ref{Remark 2.3} respectively, with $\{ \theta \in \mathds{R} | \theta \neq 1 \}$, in (\ref{(2.10-1)}) and, asymptotically  equivalently, in (\ref{(2.16-1)}), the linear combination of the respective  empirical distributions as in (\ref{(1.1)}) and (\ref{(1.5)}), $\{F_{m_{N},N}(x)-\theta F_{N}(x) , ~ x\in \mathbb{R}\}$, estimates $(1-\alpha) F(x)$ uniformly in $x \mathds{R}$. For an elaboration on this topic, we refer to \ref{Appendix}. Appendix.

\section{Numerical illustrations }\label{Numerical Illustrations}
In this section  we provide a brief numerical illustration  of conclusions  in Sections \ref{Confidence bands super population} and \ref{Confidence bands large sample} on capturing the empirical and theoretical distributions  via the sup-norm functional.

%\vspace{5.3 cm}
\begin{figure}[htb!]\label{Figure 3}
\includegraphics[width= 16 cm, height = 6.5 cm]{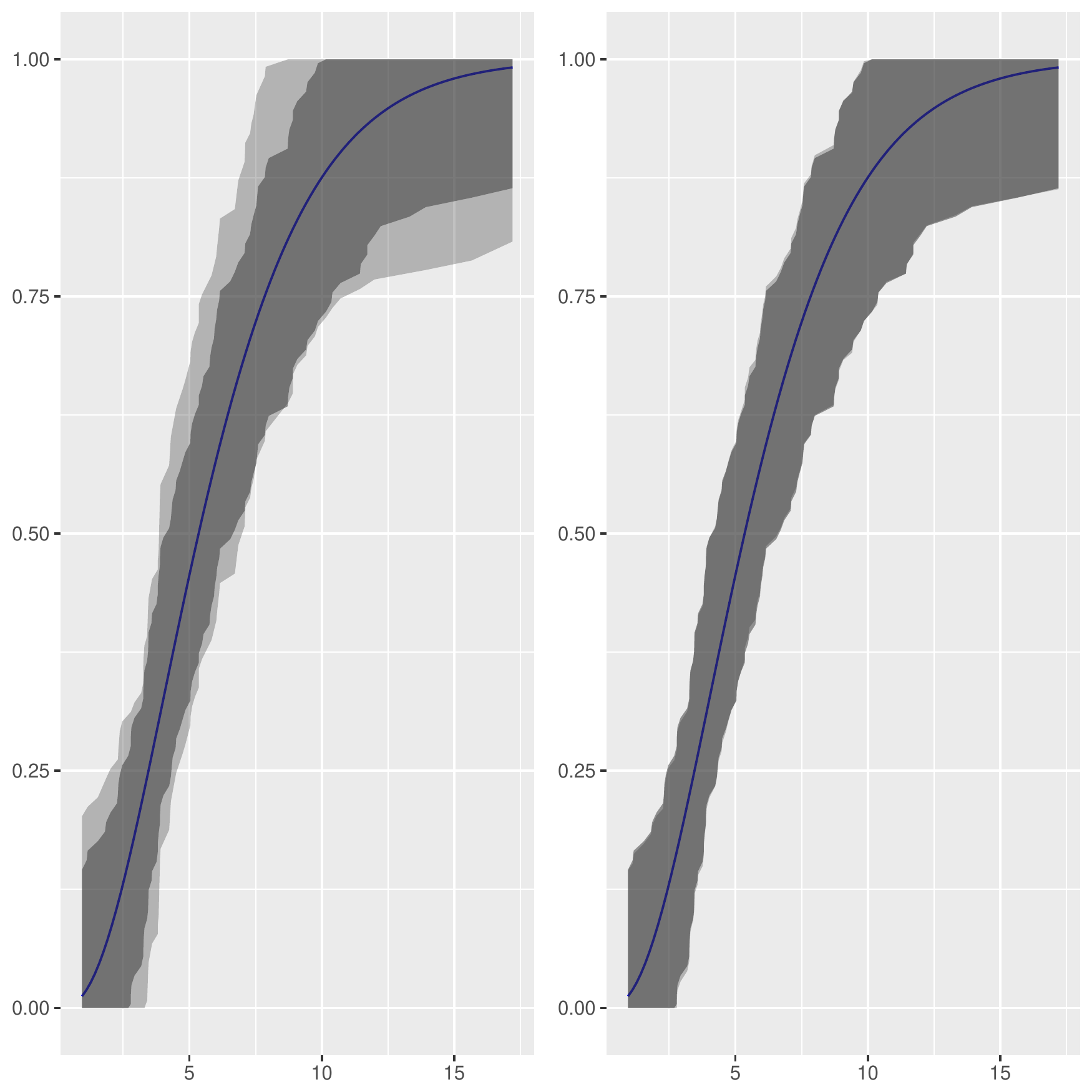}
%\vspace{-.1  cm}
\centering
\caption{ \footnotesize  Illustration of the subsampled confidence band  (\ref{(3.9)}) for $F$ in the context of Section \ref{Confidence bands large sample}. The lighter shaded subsampled confidence band (\ref{(3.9)})  overlays  the classical one (the darker shaded band)  for $F$  as in (\ref{(4.1)}),  for $F=\chi^2$ with $df=6$. The solid curve is the cdf of $F $   from which a sample of size $N=100$ was simulated. The left and right panels are 95\% confidence bands for $F$, with $c_{\alpha}=1.358$, based on subsamples of respective sizes   $m_N=N=   100$ and $m_N=   \lceil (100)^{1.9}\rceil= 6310$. The subsampled band for $F$ as in (\ref{(3.9)}) on the right   panel is just about the same as  that of  the classical one as in (\ref{(4.1)}).
   }
\end{figure}

\begin{figure}[htb!]\label{Figure 4}
%\vspace{-1 cm}
\includegraphics[width= 16 cm, height = 6.8 cm]{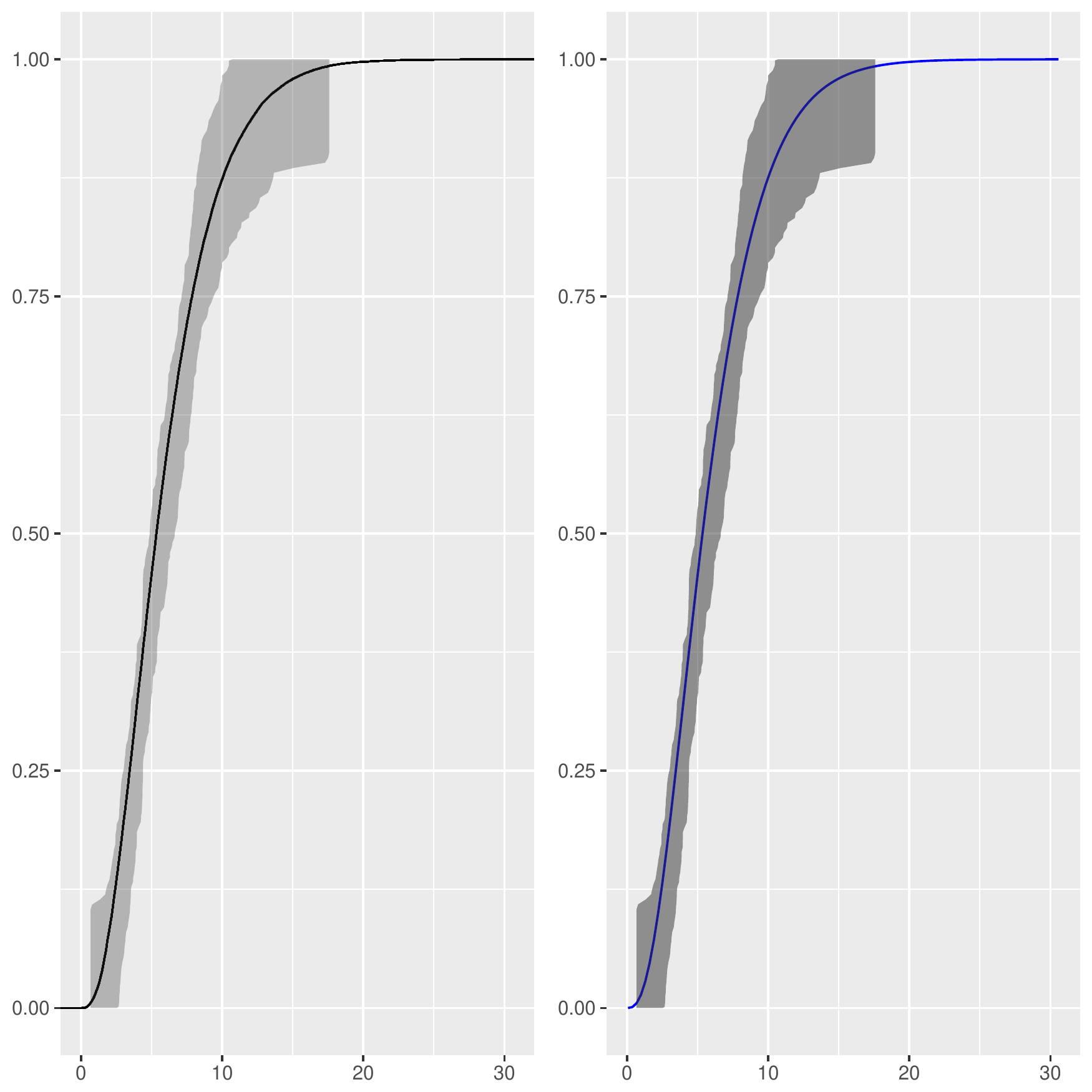}
%\vspace{-1  cm}
\centering
\caption{ \footnotesize Illustration of the subsampled confidence bands (\ref{(3.5)}) and (\ref{(3.9)}) respectively for $F_N$ and $F$ in the context of Section \ref{Confidence bands super population}. The left panel is a 95\% subsample confidence band for $F_N$ via (\ref{(3.5)}) with $c_{\alpha}=1.358$  based on a subsample of size $m_N=(36000)^{1/2}=6000$ from  a random sample of size $N=36000$ from $F=\chi^2_{6}$. The right panel is a 95\%  subsample  confidence band for $F$ via (\ref{(3.9)}) with $c_{\alpha}=1.358$ based on the same subsample of size $m_N=6000$ from  the same random sample of size    $N=36000$ from $F=\chi^2_{6}$.
}
\end{figure}

\newpage
\begin{remark}
As $N,m_N \rightarrow \infty$ so that $m_N=o(N^2)$, the thus obtained asymptotically correct band of (\ref{(3.9)}) for $F$ is a bit wider than that of (\ref{(3.5)}) for $F_N$. It, i.e., (\ref{(3.9)}), correctly views and parameterizes  the problem of estimating $F$ via the randomly weighted sample   distribution function $F_{m_N,N}$ (cf. (\ref{(1.5)})) as a two sample problem in terms of the respective subsample and sample sizes $m_N$ and $N$. On the other hand, (\ref{(3.5)}) quite naturally treats and parameterizes the problem of estimating $F_N$ via $F_{m_N,N}$ as a one sample problem in terms of having $m_N$ observations out of $N$ available for performing this inference for $F_N$. On choosing $m_N$  large so that $m_N=o(N)$ when $N$ is  large, (\ref{(3.9)}) and (\ref{(3.5)}) tend to coincide, as,   e.g., in Figure 2. However, to begin with, for the sake of establishing exact asymptotic correctness, one should   use (\ref{(3.5)}) for estimating $F_N$ and, respectively, (\ref{(3.9)}) for estimating  $F$ and, depending on how big a size $N$ one has for a concrete data set or for an imaginary random sample, one should  explore selecting smaller  values for $m_N=o(N^2)$ accordingly. For example, having a not very  large sample size $N$, say $N=10000$, one may like to consider taking $m_N=O(N)$, say $m_N=N/2$.
\end{remark}

\begin{table}[H]
\vspace{-.5 cm}
\caption{Illustration of the performance of the randomized bands (\ref{(3.9)}) for the theoretical distribution $F$ to the classical band as in (\ref{(4.1)})  based on 1000 replications of the  specified distributions. For each simulated  sample of size $N$  simultaneous subsamples  of size $m_N=\lceil{N^{a}}\rceil$, $a=0.5,  1,  1.9$, were   drawn from it to construct the  randomized  confidence bands (\ref{(3.9)}). Nominal coverage probability for the bands  is $95\%$.
}
\begin{center}
\small\addtolength{\tabcolsep}{-3pt}
\begin{tabular}{c|c|c|c c| c}
\hline
Distribution & $N$ & $m_N$  &   empiric coverage of (\ref{(3.9)}) & & empiric coverage of (\ref{(4.1)})  \\
\hline

\multirow{2}{*}{$\chi_{1}^2$}{} & \multirow{3}{*}{50}  & $ \lceil{N^{0.5}}\rceil$ & \multirow{1}{*}{0.985} & &   \\  &
                                                       &$ N$ & \multirow{1}{*}{0.976}  & & 0.970\\ &
                                                       &$ \lceil{N^{1.9}}\rceil$& \multirow{1}{*}{0.966}       & &   \\ \cline{2-6}

                                &\multirow{3}{*}{200}  & $ \lceil{N^{0.5}}\rceil$ & \multirow{1}{*}{0.978} & &   \\  &
                                                       &$ N$ & \multirow{1}{*}{ 0.971}  & & 0.953 \\ &
                                                       &$ \lceil{N^{1.9}}\rceil$& \multirow{1}{*}{0.954}       & &   \\

\hline\hline

\multirow{2}{*}{$   t_{15}$}{} & \multirow{3}{*}{50}  & $ \lceil{N^{0.5}}\rceil$ & \multirow{1}{*}{0.985} & &   \\  &
                                                       &$ N$ & \multirow{1}{*}{0.977}  & & 0.969\\ &
                                                       &$ \lceil{N^{1.9}}\rceil$& \multirow{1}{*}{0.97}       & &   \\ \cline{2-6}
                                &\multirow{3}{*}{200}  & $ \lceil{N^{0.5}}\rceil$ & \multirow{1}{*}{ 0.981} & &   \\  &
                                                       &$ N$ & \multirow{1}{*}{ 0.964}  & & 0.961 \\ &
                                                       &$ \lceil{N^{1.9}}\rceil$& \multirow{1}{*}{0.96}       & &   \\

\hline

\end{tabular}
\end{center}
%\vspace{1 cm}
\end{table}

\begin{table}[H]
\vspace{-0.5 cm}
\caption{Illustration of the performance of the randomized bands (\ref{(3.5)}) for the empirical  distribution $F_N$ based on 1000 replications.  For each simulated  sample of size $N$  simultaneous subsamples  of size $m_N=\lceil{N^{a}}\rceil$, $a=0.5, ~ 1, ~ 1.9$, were  drawn from it to construct the randomized confidence bands (\ref{(3.5)}). The nominal coverage probability for the bands is $95\%$.
}
\begin{center}
\small\addtolength{\tabcolsep}{-3pt}
\begin{tabular}{c|c|c|c c }
\hline
Distribution & $N$ & $m_N$  &  empiric  coverage of (\ref{(3.5)})     \\
\hline

\multirow{2}{*}{$\chi_{1}^2$}{} & \multirow{3}{*}{50}  & $ \lceil{N^{0.5}}\rceil$ & 0.985  &   \\  &
                                                       &$ N$ &  0.971  &      \\ &
                                                       &$ \lceil{N^{1.9}}\rceil$& 0.957 &   \\ \cline{2-5}
                                &\multirow{3}{*}{200}  & $ \lceil{N^{0.5}}\rceil$ & 0.978  &   \\  &
                                                       &$ N$ &  0.966  &      \\ &
                                                       &$ \lceil{N^{1.9}}\rceil$& 0.955  &   \\

\hline\hline

\multirow{2}{*}{$t_{15}$}{} & \multirow{3}{*}{50}  & $ \lceil{N^{0.5}}\rceil$ & 0.981  &   \\  &
                                                       &$ N$ &  0.973  &      \\ &
                                                       &$ \lceil{N^{1.9}}\rceil$& 0.967 &   \\ \cline{2-5}
                                &\multirow{3}{*}{200}  & $ \lceil{N^{0.5}}\rceil$ & 0.977  &   \\  &
                                                       &$ N$ &  0.97  &      \\ &
                                                       &$ \lceil{N^{1.9}}\rceil$& 0.954  &   \\

\hline

\end{tabular}
\end{center}
%\label{tab:multico2}
\end{table}

\section{Randomized central limit theorems and confidence intervals  for empirical and theoretical distributions at fixed points  via virtual resampling} \label{CLT}
In case of a continuous distribution function $F(.)$, our discussion  of confidence bands as in Sections \ref{Confidence bands super population} and \ref{Confidence bands large sample} can naturally be reformulated  in terms of confidence intervals, pointwise  for any fixed $x \in \IR$. In the latter context, i.e., pointwise, we can, however, do better  in general, namely also when having random samples $X_1, \ldots,X_N$, $N\geq 1$, on $X$ with an arbitrary         distribution function $F(.)$.
%For details we refer to Section 6 of Cs\"{o}rg\H{o} and Nasari (2015)

\par
As before, let $F_N (.)$ be the empirical  distribution function as in (\ref{(1.1)}). Define the standardized empirical process

\begin{eqnarray}
\alpha_N(x) &:=& \frac{N^{-1/2} \sum_{i=1}^N \Big(\ \one(X_i \leq x)-F(x) \Big)}{\sqrt{F(x)(1-F(x))}}\nonumber\\
&=& \frac{N^{1/2} \Big(F_N (x)-F(x) \Big) }{\sqrt{F(x)(1-F(x))}},~ x\in\IR,    \label{(5.1)}
\end{eqnarray}
and the sample variance of the indicator random variables $\one(X_i \leq x)$, $i=1,\ldots,N$, as

\begin{eqnarray}
S_{N}^2(x) &:=& \frac{1}{N} \sum_{i=1}^N \Big(\one(X_i \leq x)- F_N (x)   \Big)^2\nonumber\\
&=& F_N (x) (1-F_N(x)), ~ x \in \IR, \label{(5.2)}
\end{eqnarray}
where $F_N(.)$ is the empirical  distribution function.

\par
Define also the Studentized empirical process

\begin{equation}\label{(5.3)}
\hat{\alpha}_N(x) :=  \frac{N^{1/2} \Big(F_N (x)-F(x) \Big) }{\sqrt{F_N(x)(1-F_N(x))}},~ x\in\IR.
\end{equation}

\par
As a consequence of the classical central limit theorem (CLT) for Bernoulli random variables, as $N \to \infty$, we have for {\sl{ any fixed $x\in \IR$}}

\begin{equation}\label{(5.4)}
\alpha_N(x) \dto Z,
\end{equation}
where $Z$ here,  and also throughout, stands for a standard normal random variable.

\par
On combining the latter conclusion with the Glivenko-Cantelli theorem,  as $N \to \infty$, we also have

\begin{equation}\label{(5.5)}
\hat{\alpha}_N(x) \dto Z
\end{equation}
for {\textsl{any fixed}} $x\in \IR$.

\par
With $m_N=\sumN w_{i}^{(N)}$ and multinomial weights $\Big(w_{1}^{(N)},\ldots,w_{N}^{(N)}  \Big)$ as in (\ref{(1.3)}) that are independent from the random sample $X_1,\ldots,X_N$, define the {\sl{randomized}} standardized empirical process $\alpha_{m_N,N}^{(1)} (x)$,  a standardized version of $\beta_{m_N,N}^{(1)}(x)$ as in (\ref{(1.4)}), as follows

\begin{eqnarray}
\alpha_{m_N,N}^{(1)} (x) &:=& \frac{\sumN \big( \frac{w_{i}^{(N)}}{m_N} -\frac{1}{N} \big) \one(X_i \leq x)}{\sqrt{F(x)(1-F(x))} \sqrt{\sum_{j=1}^N \big( \frac{w_{j}^{(N)}}{m_N} -\frac{1}{N} \big)^2}}\nonumber\\
&=& \frac{F_{m_N,N}(x)-F_N(x)}{\sqrt{F(x)(1-F(x))} \sqrt{\sum_{j=1}^N \big( \frac{w_{j}^{(N)}}{m_N} -\frac{1}{N} \big)^2}}, ~ x \in \IR, \label{(5.6)}
\end{eqnarray}
where $F_{m_N,N}(.)$ is the randomly weighted sample distribution function as in (\ref{(1.5)}).

\par
Define also the randomized subsample variance of the indicator random variables $\one(X_i \leq x)$, $i=1,\ldots,N$, as

\begin{eqnarray}
S_{m_N,N}^2 &:=&\frac{1}{m_N} \sumN w_{i}^{(N)} \Big(  \one(X_i \leq x) -F_{m_N,N} \Big)^2 \nonumber \\
&=&  F_{m_N,N}(x) (1-F_{m_N,N}(x)), ~ x \in \IR. \label{(5.7)}
\end{eqnarray}

\par
With $N$ fixed and $m=m_N \to \infty$, via (\ref{(1.13)}),   pointwise in $x \in \IR$,  we arrive at

\begin{equation}\label{(5.8)}
S_{m_N,N}^{2}(x) \to F_N (x) (1-F_N(x))=S_{N}^{2}(x)~ in ~ probability~ P_{X,w},
\end{equation}
and, as a consequence of (\ref{(1.14)}), as $N,m_N \to \infty$, pointwise in $x \in \IR$, we also  have

\begin{equation}\label{(5.9)}
\Big(S_{m_N,N}^2(x) -S_{N}^{2}(x)\Big) \to 0~ in ~ probability~ P_{X,w}
\end{equation}
with $S_{m_N,N}^2(x)$ and $S_N^{2}(x)$ respectively as in (\ref{(5.7)}) and (\ref{(5.2)}).

\par
Further to the randomized standardized empirical process $\alpha^{(1)}_{m_N,N}(x)$, we now define

\begin{eqnarray}
\alpha_{m_N,N}^{(2)} (x) &:=& \frac{\sum_{i=1}^N  \frac{w_{i}^{(N)}}{m_N}  \Big(\one(X_i \leq x) -F(x) \Big)}{\sqrt{F(x)(1-F(x))} \sqrt{\sum_{j=1}^N \big( \frac{w_{j}^{(N)}}{m_N}  \big)^2}}\nonumber\\
&=& \frac{F_{m_N,N}(x)-F(x)}{\sqrt{F(x)(1-F(x))} \sqrt{\sum_{j=1}^N \big( \frac{w_{j}^{(N)}}{m_N} -\frac{1}{N} \big)^2 +\frac{1}{N}}}, ~ x \in \IR, \label{(5.10)}
\end{eqnarray}
a standardized version of $\beta_{m_N,N}^{(2)}(x)$ as in (\ref{(1.6)}) with an  arbitrary distribution function $F(.)$.

\par
Along the above lines, we also define the standardized version of $\beta_{m_N,N}^{(3)}(x, \theta)$ of (\ref{(1.7)}), with an arbitrary distribution function $F(.)$, namely

\begin{eqnarray}
\alpha^{(3)}_{m_N,N}(x,\theta) &:=& \frac{\sumN \Big( \frac{w_{i}^{(N)}}{m_N} - \frac{\theta}{N} \Big) \Big( \one(X_i \leq x) - F(x) \Big) }{\sqrt{F(x)(1-F(x))} \sqrt{\sum_{j=1}^N \Big( \frac{w_{i}^{(N)}}{m_N} - \frac{\theta}{N} \Big)^2} } \nonumber\\
&=&\frac{\Big(F_{m_N,N}(x)-F(x)\Big)-\theta\Big(F_N(x)-F(x) \Big) }{ \sqrt{F(x)(1-F(x))} \sqrt{\sum_{j=1}^N \Big( \frac{w_{j}^{(N)}}{m_N}- \frac{1}{N} \Big)^2 + \frac{(1-\theta)^2}{N} } }, ~~ x \in \IR, \label{(5.11)}
\end{eqnarray}
where $\theta$ is a real valued constant.

\begin{remark}\label{Remark 5.1}
On letting $\theta=1$ in (\ref{(5.11)}), it reduces to $\alpha^{(1)}_{m_N,N}(x)$ of (\ref{(5.6)}), while letting $\theta=0$ in (\ref{(5.11)}) yields  $\alpha^{(2)}_{m_N,N}(x)$ of (\ref{(5.10)}).
\end{remark}

\par
For the  proof of the  results of our next proposition, we refer to Remark \ref{Remark 6.2}.

\begin{prop}\label{Proposition 5.1}
As $N,m_N \to \infty$ so that $m_N=o(N^2)$,  for  $\alpha^{(3)}_{m_N,N}(.,\theta)$ as in (\ref{(5.11)}),  we have the following central limit theorem (CLT)

\begin{equation}\label{(5.12)}
P_{X|w} \Big( \alpha^{(3)}_{m_N,N}(x,\theta) \leq t  \Big) \longrightarrow \Phi(t)~ in ~ probability~ P_w
\end{equation}
for all $x,t \in \IR$, as well as the unconditional CLT

\begin{equation}\label{(5.13)}
P_{X,w} \Big( \alpha^{(3)}_{m_N,N}(x,\theta) \leq t  \Big) \longrightarrow \Phi(t)~ for ~all ~ x,t \in \IR,
\end{equation}
where $\Phi(.)$ is the unit normal distribution function.

\end{prop}

\begin{corollary}\label{Corollary 5.1}
In view of Remark \ref{Remark 5.1}  and Proposition \ref{Proposition 5.1}, as $N,m_N \to \infty$ so that $m_N=o(N^2)$, we have

\begin{equation}\label{(5.14)}
\alpha^{(s)}_{m_N,N}(x) \disto Z ~~ for ~all ~ x \in \IR,
\end{equation}
with $s=1$ (cf. (\ref{(5.6)})) and also for $s=2$ (cf. (\ref{(5.10)})), both CLTs in terms of both $P_{X|w}$ (cf. (\ref{(5.12)}))  and $P_{X,w}$ (cf. (\ref{(5.13)})).
\end{corollary}

\begin{remark}\label{Remark 5.2}
On combining the two respective  conclusions of Proposition \ref{Proposition 5.1} with the Glivenko-Cantelli theorem, the latter two continue to hold true with $\sqrt{F_N(x)  (1-F_N(x))}$ replacing $\sqrt{F(x)  (1-F(x))}$ in the definition of $\alpha_{m_N,N}^{(3)}(x,\theta)$ as in (\ref{(5.11)}), as well as with $\sqrt{F_{m_N,N}(x)  (1-F_{m_N,N}(x))}$ replacing $\sqrt{F(x)  (1-F(x))}$ therein,  on account of the conclusion of (\ref{(5.9)}). Consequently, similar respective versions of (\ref{(5.14)}) of Corollary \ref{Corollary 5.1} also hold true. We are now to spell out these conclusions  in our next three corollaries.
\end{remark}

\begin{corollary}\label{Corollary 5.2-1}
As $N,m_N \to \infty$ so that $m_N=o(N^2)$, we have in terms of both $P_{X|w}$ and $P_{X,w}$
\begin{eqnarray}\label{(5.15-1)}
\hat{\alpha}_{m_N,N}^{(3)}(x,\theta) &:=&  \frac{\big( F_{m_N,N}(x) - \theta F_N(x)\big)- \big(1-\theta \big) F(x) }{\sqrt{F_{N}(x) (1-F_{N}(x)) }  \sqrt{\sum_{j=1}^N  ( \frac{w_{j}^{(N)}}{m_N}-\frac{1}{N} )^2 +\frac{(1-\theta)^2}{N}}} \nonumber \\
&& \disto Z ~ for ~ all~ x\in \IR,
\end{eqnarray}
and

\begin{eqnarray}\label{(5.16-1)}
\hat{\hat{\alpha}}_{m_N,N}^{(1)}(x,\theta) &:=&  \frac{\big( F_{m_N,N}(x) - \theta F_N(x)\big)- \big(1-\theta \big) F(x) }{\sqrt{F_{m_N,N}(x) (1-F_{m_N,N}(x)) }  \sqrt{\sum_{j=1}^N  ( \frac{w_{j}^{(N)}}{m_N}-\frac{1}{N} )^2 +\frac{(1-\theta)^2}{N}}}\nonumber \\
 && \disto Z ~ for ~ all~ x\in \IR,
\end{eqnarray}
with any constant $\theta \in \IR$ in both cases.
\end{corollary}

\par
On taking $\theta=1$, respectively $\theta=0$, in Corollary \ref{Corollary 5.2-1}, we arrive at the following two corollaries.

\begin{corollary}\label{Corollary 5.2}
As $N,m_N \to \infty$ so that $m_N=o(N^2)$, we have in terms of both $P_{X|w}$ and $P_{X,w}$

\begin{eqnarray}\label{(5.15)}
\hat{\alpha}_{m_N,N}^{(1)}(x) &:=& \hat{\alpha}_{m_N,N}^{(3)}(x, 1)= \frac{F_{m_N,N}(x) - F_N(x) }{\sqrt{F_{N}(x) (1-F_{N}(x)) }  \sqrt{\sum_{j=1}^N \Big( \frac{w_{j}^{(N)}}{m_N}-\frac{1}{N} \Big)^2 }} \nonumber \\
& &\disto Z ~ for ~ all~ x\in \IR,
\end{eqnarray}
and

\begin{eqnarray}\label{(5.16)}
\hat{\hat{\alpha}}_{m_N,N}^{(1)}(x) := \hat{\hat{\alpha}}_{m_N,N}^{(3)}(x,1) &=& \frac{F_{m_N,N}(x) - F_N(x) }{\sqrt{F_{m_N,N}(x) (1-F_{m_N,N}(x)) }  \sqrt{\sum_{j=1}^N \Big( \frac{w_{j}^{(N)}}{m_N}-\frac{1}{N} \Big)^2 }} \nonumber \\
&&\disto Z ~ for ~ all~ x\in \IR.
\end{eqnarray}

\end{corollary}

\begin{corollary}\label{Corollary 5.3}
As $N,m_N \to \infty$ so that $m_N=o(N^2)$, we have in terms of both $P_{X|w}$ and $P_{X,w}$

\begin{eqnarray}\label{(5.17)}
\hat{\alpha}_{m_N,N}^{(2)}(x) := \hat{\alpha}_{m_N,N}^{(3)}(x,0) &=&\frac{F_{m_N,N}(x) - F(x) }{\sqrt{F_{N}(x) (1-F_{N}(x)) }  \sqrt{\sum_{j=1}^N \Big( \frac{w_{j}^{(N)}}{m_N}-\frac{1}{N} \Big)^2 + \frac{1}{N} }}\nonumber\\
&\disto& Z ~ for ~ all~ x\in \IR,
\end{eqnarray}
and

\begin{eqnarray}\label{(5.18)}
\hat{\hat{\alpha}}_{m_N,N}^{(2)}(x) := \hat{\hat{\alpha}}_{m_N,N}^{(3)}(x,0) &=& \frac{F_{m_N,N}(x) - F(x) }{\sqrt{F_{m_N,N}(x) (1-F_{m_N,N}(x)) }  \sqrt{\sum_{j=1}^N \Big( \frac{w_{j}^{(N)}}{m_N}-\frac{1}{N} \Big)^2 +\frac{1}{N} }} \nonumber\\
& \disto& Z ~ for ~ all~ x\in \IR.
\end{eqnarray}

\end{corollary}

\par
In view of  Lemma \ref{Lemma 6.2},  Corollaries \ref{Corollary 5.2-1}, \ref{Corollary 5.2}  and \ref{Corollary 5.3} have the following asymptotically equivalent respective  forms with more familiar looking norming sequences  that are also more convenient for doing calculations.

\begin{corollary}\label{Corollary 5.4-1}
As $N,m_N \to \infty$ so that $m_N=o(N^2)$, we have in terms of both $P_{X|w}$ and $P_{X,w}$

\begin{eqnarray}\label{(5.19-1)}
\tilde{\alpha}_{m_N,N}^{(3)}(x,\theta) &:=&  \sqrt{\frac{N m_N}{N+m_N(1-\theta)^2} }
\frac{\big(F_{m_N,N}(x) - \theta F_N(x)\big)-\big( 1-\theta \big) F(x) }{\sqrt{F_{N}(x)
(1-F_{N}(x)) }  } \nonumber \\
&\disto& Z ~ for ~ all~ x\in \IR,
\end{eqnarray}
and

\begin{eqnarray}\label{(5.20-1)}
\tilde{\tilde{\alpha}}_{m_N,N}^{(3)}(x,\theta) &:=&  \sqrt{\frac{N m_N}{N+m_N(1-\theta)^2} }
\frac{\big(F_{m_N,N}(x) - \theta F_N(x)\big)-\big( 1-\theta \big) F(x) }{\sqrt{F_{m_N,N}(x)
(1-F_{m_N,N}(x)) }  } \nonumber \\
&\disto& Z ~ for ~ all~ x\in \IR,
\end{eqnarray}
with any constant $\theta \in \IR$ in both cases.
\end{corollary}

\par
On taking $\theta=1$, respectively $\theta=0$, in Corollary \ref{Corollary 5.4-1}, we
arrive at the following two corollaries, respective companions of Corollaries \ref{Corollary 5.2}
and \ref{Corollary 5.3}.

\begin{corollary}\label{Corollary 5.4}
As $N,m_N \to \infty$ so that $m_N=o(N^2)$, we have in terms of both $P_{X|w}$ and $P_{X,w}$

\begin{equation}\label{(5.19)}
\tilde{\alpha}_{m_N,N}^{(1)}(x) := \tilde{\alpha}_{m_N,N}^{(3)}(x,1) = \frac{\sqrt{m_N}\Big(F_{m_N,N}(x) - F_N(x)\Big) }{\sqrt{F_{N}(x) (1-F_{N}(x)) }  } \disto Z ~ for ~ all~ x\in \IR,
\end{equation}
and

\begin{equation}\label{(5.20)}
\tilde{\tilde{\alpha}}_{m_N,N}^{(1)}(x) := \tilde{\tilde{\alpha}}_{m_N,N}^{(3)}(x,1)= \frac{\sqrt{m_N}\Big(F_{m_N,N}(x) - F_N(x)\Big) } {\sqrt{F_{m_N,N}(x) (1-F_{m_N,N}(x))} } \disto Z ~ for ~ all~ x\in \IR.
\end{equation}

\end{corollary}

\begin{corollary}\label{Corollary 5.5}
As $N,m_N \to \infty$ so that $m_N=o(N^2)$, we have in terms of both $P_{X|w}$ and $P_{X,w}$

\begin{equation}\label{(5.21)}
\tilde{\alpha}_{m_N,N}^{(2)}(x) := \tilde{\alpha}_{m_N,N}^{(3)}(x,0)= \sqrt{\frac{N m_N}{N+m_N}} \frac{\Big(F_{m_N,N}(x) - F(x) \Big)}{ \sqrt{F_{N}(x) (1-F_{N}(x)) }}   \disto Z ~ for ~ all~ x\in \IR,
\end{equation}
and

\begin{equation}\label{(5.22)}
\tilde{\tilde{\alpha}}_{m_N,N}^{(2)}(x) := \tilde{\tilde{\alpha}}_{m_N,N}^{(3)}(x,0) = \sqrt{\frac{N m_N}{N+m_N}} \frac{F_{m_N,N}(x) - F(x)}{     \sqrt{F_{m_N,N}(x) (1-F_{m_N,N}(x)) } }   \disto Z ~ for ~ all~ x\in \IR.
\end{equation}

\end{corollary}

\par
In the context of viewing \textit{big data sets} and \textit{finite populations}  as random samples and imaginary random samples respectively from an infinite super-population, it is of interest to estimate both $F_N(x)$ and $F(x)$ also \textit{pointwise} in $x\in \mathbb{R}$. The asymptotically equivalent CLT's for $\hat{\hat{\alpha}}^{(1)}_{m_N,N} (x)$ and $\tilde{\tilde{\alpha}}^{(1)}_{m_N,N} (x)$ as in (\ref{(5.16)}) and (\ref{(5.20)}) can be used to construct asymptotically exact $(1-\alpha)$ size confidence sets for any $\alpha \in (0,1)$ and pointwise  in $x \in \mathbb{R}$ for the empirical distribution function $F_N (x)$ in terms of both $P_{X|w}$ and  $P_{X,w}$.
We note in passing that these two CLT's are essential extensions of the pointwise in $x \in \IR$  estimation of $F_N (x)$ by $F_{m_N,N}(x)$ in (\ref{(1.14)}), when $N, m_N \to \infty$ so that $m_N=o(N^2)$ is assumed.

\par
In the same context, the asymptotically  equivalent CLT's for $\hat{\hat{\alpha}}^{(2)}_{m_N,N} (x)$ and $\tilde{\tilde{\alpha}}^{(2)}_{m_N,N} (x)$ as in (\ref{(5.18)}) and (\ref{(5.22)}) can, in turn, be used to construct asymptotically exact $(1-\alpha)$ size confidence intervals for any $\alpha \in (0,1)$ and pointwise in $x \in \mathbb{R}$ for the arbitrary distribution function $F(x)$, in terms of both $P_{X|w}$ and $P_{X,w}$. We remark as well that these two CLT's constitute significant extensions of the poinwise in $x \in \mathbb{R}$ estimation of $F(x) $ by $F_{m_N,N}(x)$ in view of (\ref{(1.14)}), when $N,m_N \to \infty$ so that $m_N=o(N^2)$ is assumed.

\par
As to how to go about constructing these confidence intervals in hand, and to illustrate the kind of reduction of the number of data we can work with in context of a big data set or a finite population, say of of size $N=10^4$, mutatis mutandis, we refer back to the illustrative example  right after (\ref{(3.5)}), where  we outline taking  virtual sub-samples of sizes $m_{10^4}= \sqrt{10^4}=100$.

\par
The CLT's  for  ${\hat{\alpha}}^{(1)}_{m_N,N} (.)$ and  $\hat{\hat{\alpha}}^{(1)}_{m_N,N} (.)$ in Corollary \ref{Corollary 5.2} were already concluded on their own in Cs\"{o}rg\H{o} and Nasari (2015) (cf. the respective conclusions of (62) and (63) with $s=1$ and that of (64) in Section 6 therein), where the use of $\hat{\hat{\alpha}}^{(1)}_{m_N,N} (.)$ for constructing pointwise confidence sets for $F_N (.)$ and pointwise confidence intervals for $F(.)$ is also detailed.

\par
When \textit{all} $N$ \textit{observables} are \textit{available}  and desirable to be processed, then it is inviting to study and compare  the asymptotic confidence intervals that are respectively provided for the arbitrary distribution function $F(x)$ pointwise in $x \in \mathbb{R}$ by the classical Studentized process as in (\ref{(5.3)}), via (\ref{(5.5)}), and the ones we can construct using the randomized Studentized empirical processes, say via the CLT's as in (\ref{(5.21)}) and (\ref{(5.22)}) respectively, both with $m_N=N$,  and $N \to \infty$. Also, more generally, when indexed by $\{\theta \in \IR|~ \theta \neq 1 \}$, the CLT's in (\ref{(5.19-1)}) and (\ref{(5.20-1)}) yield a family of confidence intervals for an arbitrary distribution function $F(x)$ pointwise in $x \in \IR$ that,   as $N, m_N \to \infty$  so that $m_N=o(N^2)$, can be studied along the lines of Section \ref{Confidence bands large sample} and \ref{Appendix}. Appendix.

\section{Proofs }\label{Proofs}
Our Theorem \ref{THEOREM} for the randomly weighted empirical processes, respectively   as in (\ref{(1.7)}) and (\ref{(1.8)})  that,  with $F(x)$ assumed to be continuous,   read as

\begin{equation}\nonumber
\{\beta_{m_N,N}^{(3)}(x,\theta), x \in \IR \}= \{\beta_{m_N,N}^{(3)}(F^{-1}(t),\theta), 0\leq t \leq 1 \}
\end{equation}
and
\begin{equation}\nonumber
\{\tilde{\beta}_{m_N,N}^{(3)}(x,\theta), x \in \IR \}= \{\tilde{\beta}_{m_N,N}^{(3)}(F^{-1}(t),\theta), 0\leq t \leq 1 \},
\end{equation}
is based on well known results on the weak convergence of scalar-weighted empirical processes (cf., e.g., Section 3.3 in Shorack and Wellner (1986) and Section 2.2 in Koul (2002)).  The original papers dealing with such empirical processes date back to Koul (1970) and Koul and Staudte (1972) (cf.\ Shorack (1979) for further  references  in this regard).  In our context, we make use of Corollary 2.2.2 in Koul (2002), and spell it out as follows.

\begin{thm}\label{Theorem 6.1}
%variables on $(\Omega_X,\calF_X,P_X)$ and assume that the distribution function $F(x)=P_X(X\le x)$ is continuous.  Let $\Big\{d_{i,N}\Big\}^N_{i=1}$,  $N\ge 1$, be a triangular array of real numbers, and define the weighted empirical process
Let $X,X_1, \ldots$ be real valued i.i.d. random variables on $(\Omega_X, \mathfrak{F}_X,P_X)$ and assume that the distribution function $F(x)=P_{X}(X \leq x)$ is continuous. Let $\big\{d_{i,N}\big\}_{i=1}^{N}$ be a triangular array of real numbers, and define the weighted empirical process
\begin{eqnarray*}
\beta_{d,N}(x) &:=& \sum^N_{i=1} d_{i,N}(\one(X_i\le x)-F(x)), ~x\in\IR,\\
&=& \sum^N_{i=1} d_{i,N}(\one(F(X_i) \le t)-t)\phantom{WWW}\\
&=:& \beta_{d,N}(F^{-1}(t)), ~~0\le t\le 1. \label{(6.1)}
\end{eqnarray*}
Assume that, as $N\to\infty$,

\begin{equation}
\mathcal{H}_N := \max_{1\le i\le  N} d^2_{i,N} \to 0, \label{(6.2)}\\
\end{equation}
and
\begin{equation}
\sum^N_{i=1} d^2_{i,N} = 1 \hbox{ ~ for each ~} N\ge 1.
\label{(6.3)}
\end{equation}
Then, as $N\to\infty$,

\begin{equation}\label{(6.4)}
\beta_{d,N}(F^{-1}(t))  \lawto B(t) \hbox{ ~on~ } (D,\calD, \Vert~\Vert),
\end{equation}
where $B(.)$ is a Brownian bridge on $[0,1]$, $\calD$ denotes the $\sigma$-field generated by the finite dimensional subsets of  $D=D[0,1]$, and $\Vert ~ \Vert$ stands for the uniform metric for real valued functions on $[0,1]$.
\end{thm}
In order to ``translate'' Theorem \ref{Theorem 6.1} to the first conclusion (\ref{(2.2)}) of  our Theorem \ref{THEOREM}, it will suffice  to conclude the following maximal negligibility of the weights in probability $P_w$.

%we quote Lemma 5.2 of \Cs, Martsynyuk and Nasari (2014) as follows.

\begin{lemma}\label{Lemma 6.1}
Let $N,m_N \to \infty$ so that $m_N=o(N^2)$.  Then
\begin{equation}\label{(6.5)}
\mathcal{H}_N=\mathcal{H}_N (\theta) := \frac{\displaystyle{\max_{1\le i\le N}}\Big(
\frac{w_i^{(N)}}{m_N} - \frac{\theta}{N}\Big)^2}
{
\displaystyle{\sum^N_{j=1}}  \Big(  \frac
{w_j^{(N)}}{m_N} - \frac{\theta}{N}\Big)^2 }
\to 0  \hbox{\sl ~in probability}\ \!P_w
\end{equation}
with any arbitrary real constant $\theta$.
\end{lemma}

\begin{remark}\label{Remark 6.1}
With $\theta=1$, the conclusion (\ref{(6.5)}) of Lemma \ref{Lemma 6.1} was established in Cs\"{o}rg\H{o}, Martsynyuk and Nasari (2014) in their Lemma 5.2.
\end{remark}

\par
The second conclusion (\ref{(2.3)}) of Theorem \ref{THEOREM} follows from that of its (\ref{(2.2)}) via the following asymptotic equivalence of the weights in probability $P_w$.

\begin{lemma}\label{Lemma 6.2}
 Let $N,m_N \to\infty$ so that $m_N=o(N^2)$. Then
%in  addition to (\ref{(6.5)}), we need, and have as well,\\
\begin{equation}\label{(6.6)}
\left| \sum^N_{i=1}\left( \frac{w_i^{(N)}}{m_N} - \frac{1}{N} \right)^2 - \frac{1}{m_N} \right| = o_{P_w}(1).
\end{equation}
\end{lemma}
%Theorem \ref{THEOREM A} combined with Lemma \ref{LEMMA} imply that, as $m_N,N \to \infty$ such that $m_N=o(N^2)$,  $\beta_{m_N,N}^{(1)}$   converge weakly to $B(.)$ on $(D,\calD, \Vert~\Vert)$. The same also hold true for $\beta_{m_N,N}^{(3)}$, in view of (\ref{(6.6)}).
%\\
%To complete the proof of this theorem it remains to show that, as $m_N,N \to \infty$ such that $m_N=o(N)$, $\beta_{m_N,N}^{(t)}$, $s=2,4$, also  converge weakly to $B(.)$ on $(D,\calD, \Vert~\Vert)$.  To do so we first note that, for all $x \in \IR$, by virtue of (\ref{(6.6)}) we write

%\begin{eqnarray*}
%\beta_{m_N,N}^{(2)}(x) &=& \beta_{m_N,N}^{(1)}(x) + \frac{F_N(x)-F(x)}{\sqrt{\sum^N_{i=1}\left( \frac{w_i^{(N)}}{m_N} - \frac{1}{N} \right)^2}}\\
%&=& \beta_{m_N,N}^{(1)}(x)+ \beta_{N}(x)\Bigg( \frac{\big(\sqrt{m_N} +o_{P_w}(1)\big)}{\sqrt{N}} \Bigg).
%\end{eqnarray*}
Based on Lemmas \ref{Lemma 6.1} and \ref{Lemma 6.2}, we are to  first conclude Theorem \ref{THEOREM} via  Theorem \ref{Theorem 6.1}, and then proceed to prove these two lemmas in hand.

\subsection*{Proof of Theorem \ref{THEOREM}}
Put

\begin{equation}\nonumber
d_{i,N}=d_{i,N}(\theta):= \frac{w_{i}^{(n)}/m_N - \theta/N}{\sqrt{(w_{i}^{(n)}/m_N - \theta/N)^2}}, ~ 1\leq i \leq N.
\end{equation}
Clearly

\begin{equation}\label{(6.6)}
\sum_{i=1}^N d_{i,N}^2=1, ~ \textrm{for \ each }\ N\geq 1,
\end{equation}
and, in view of Lemma \ref{Lemma 6.1}, as $N,m_N \to \infty$ so that $m_N=o(N^2)$,

\begin{equation}\label{(6.7)}
\mathcal{H}_N:= \max_{1\leq i \leq N} d^{2}_{i,N} \to \textrm{in\  probability} \ P_w.
\end{equation}
Consequently, every subsequence $\{N_k\}_k$ of $N$, $N\geq 1$, contains a further subsequence $\{N_{k_{l}}\}_{l}$ such that as $l \to \infty$,  then

\begin{equation}\label{(6.8)}
\mathcal{H}_{N_{k_{l}}}:= \max_{1\leq i \leq N_{k_{l}}} d^{2}_{i,N_{k_{l}}} \to \textrm{almost\  surely\ in}\ P_w.
\end{equation}
In view of the characterization of convergence in probability in terms of almost sure convergence of subsequences, the respective  conclusions of (\ref{(6.7)}) and (\ref{(6.8)}) are equivalent to each other,  and the latter holds true for a set of weights $\tilde{\Omega}_{w} \in \mathfrak{F}_w$ with $P(\tilde{\Omega}_{w})=1$.

\par
Thus, in our present context, condition  (\ref{(6.1)})    of   Theorem \ref{Theorem 6.1}   is satisfied almost surely in $P_w$ as in (\ref{(6.8)}),  and that via (\ref{(6.7)}) leads to concluding the statement of (\ref{(2.2)}) of Theorem \ref{THEOREM}. Also, the latter in combination with Lemma \ref{Lemma 6.2} leads to concluding the second statement (\ref{(2.3)}) of Theorem \ref{THEOREM} as well. $\square$

\begin{remark}\label{Remark 6.2}
Similarly to the above proof of (\ref{(2.2)}) of Theorem \ref{THEOREM}, via the Lindeberg-Feller CLT based   Lemma 5.1 of Cs\"{o}rg\H{o} \textit{et al}. (2014) in combination with our  Lemma \ref{Lemma 6.1}, we arrive at concluding Proposition \ref{Proposition 5.1}. The latter, in turn, combined with Lemma \ref{Lemma 6.2} results in having also Corollary \ref{Corollary 5.4-1}. Thus, the content of Section \ref{CLT} becomes self-contained.
\end{remark}

\par
Using on occasions Lemma \ref{Lemma 6.2} as well, we now proceed with proving Lemma \ref{Lemma 6.1}. In turn, we then prove Lemma \ref{Lemma 6.2}.

\subsection*{Proof of Lemma \ref{Lemma 6.1}}
First we show that as $N,m_N \to \infty $ so that $m_N=o(N^2)$, then

\begin{equation}\label{(6.9)}
\frac{\max_{1\leq i \leq N} \big(\frac{w_{i}^{(N)}}{m_N}-\frac{\theta}{N}   \big)^2 }  { \sum_{j=1}^N \big(\frac{w_{j}^{(N)}}{m_N}-\frac{1}{N}   \big)^2} \longrightarrow 0 ~ \textrm{in} ~P_w
\end{equation}
with any arbitrary real constant $\theta$.
\par
We have

\begin{eqnarray}
\big(\frac{w_{i}^{(N)}}{m_N}-\frac{\theta}{N}   \big)^2 &=& \big(\frac{w_{i}^{(N)}}{m_N}-\frac{1}{N}+\frac{1}{N}-\frac{\theta}{N}   \big)^2\nonumber\\
&=& \big(\frac{w_{i}^{(N)}}{m_N}-\frac{1}{N}   \big)^2 + 2\big(\frac{w_{i}^{(N)}}{m_N}-\frac{1}{N}   \big) \big( \frac{1}{N} - \frac{\theta}{N} \big)+ \big( \frac{1}{N} - \frac{\theta}{N} \big)^2, ~ 1 \leq i \leq N. \nonumber
\end{eqnarray}
Consequently,

\begin{equation}\label{(6.10)}
\frac{\max_{1\leq i \leq N} \big(\frac{w_{i}^{(N)}}{m_N}-\frac{\theta}{N}   \big)^2 }  { \sum_{j=1}^N \big(\frac{w_{j}^{(N)}}{m_N}-\frac{1}{N}   \big)^2}  \leq \mathcal{H}_{N}(1)+I_{2}(N)+I_{3}(N),
\end{equation}
where $\mathcal{H}_{N}(1)$ is $\mathcal{H}_{N}=\mathcal{H}_{N}(\theta)$ with $\theta=1$, as in (\ref{(6.5)}), i.e., as $N,m_N \to \infty $ so that $m_N=o(N^2)$, via Remark \ref{Remark 6.1},

\begin{equation}\label{(6.11)}
\mathcal{H}_{N}(1)=o_{P_w}(1),
\end{equation}

\begin{eqnarray}
I_{2}(N) &:=& 2 |1-\theta| \big( \frac{\max_{1\leq i \leq N } |\frac{w_{i}^{(N)}}{m_N} - \frac{1}{N} | }{\sqrt{\sum_{j=1}^N \big( \frac{w_{j}^{(N)}}{m_N} - \frac{1}{N}   \big)^2}} \big) \big(\frac{1}{N\sqrt{\sum_{j=1}^N \big( \frac{w_{j}^{(N)}}{m_N} - \frac{1}{N}   \big)^2}} \big) \nonumber\\
&=& o_{P_w}(1) \big(\frac{1}{N\sqrt{\sum_{j=1}^N \big( \frac{w_{j}^{(N)}}{m_N} - \frac{1}{N}   \big)^2}} \big) \nonumber\\
&=& o_{P_w}(1) \big(\frac{1}{N\sqrt{o_{P_w}(1) +1/m_N }} \big) \nonumber \\
&=& o_{P_w}(1) \big(\sqrt{m_N}/N)  \big(\frac{1}{\sqrt{m_N o_{P_w}(1) +1  }} \big) \nonumber \\
&=& o_{P_w}(1) o(1)    \big(\frac{1}{\sqrt{m_N o_{P_w}(1) +1  }} \big) \nonumber \\
&=& o_{P_w}(1), \label{(6.12)}
\end{eqnarray}
in view of (\ref{(6.11)}), Lemma \ref{Lemma 6.2}, and $m_N=o(N^2)$, and

\begin{eqnarray}
I_3 (N) &:=& \frac{(1-\theta)^2}{N^2 \sum_{j=1}^N \big( \frac{w_{j}^{(N)}}{m_N} - \frac{1}{N}   \big)^2}= \frac{(1-\theta)^2}{N^2 (o_{P_w}(1)+1/m_N)} \nonumber\\
&=& (m_N/N^2) (1-\theta)^2 \frac{1}{m_N o_{P_w}(1)+1}=o(1) \frac{(1-\theta)^2}{m_N o_{P_w}(1)+1} \nonumber \\
&=&o_{P_w}(1), \label{(6.13)}
\end{eqnarray}
in view of Lemma \ref{Lemma 6.2} and $m_N=o(N^2)$.

\par
Combining now (\ref{(6.11)}) , (\ref{(6.12)}) and (\ref{(6.13)}), via (\ref{(6.10)}) we arrive at (\ref{(6.9)}).

\par
Next we write $\mathcal{H}_{N}=\mathcal{H}_{N}(\theta)$ of (\ref{(6.4)}) as

\begin{eqnarray}\label{(6.14)}
\mathcal{H}_{N}(\theta)&=& \big(\frac{\max_{1\leq i \leq N} \big(\frac{w_{i}^{(N)}}{m_N}-\frac{\theta}{N}   \big)^2      }  { \sum_{j=1}^N \big(\frac{w_{j}^{(N)}}{m_N}-\frac{1}{N}   \big)^2 } \big) \big(  \frac{ \sum_{j=1}^N \big(\frac{w_{j}^{(N)}}{m_N}-\frac{1}{N}   \big)^2  }{\sum_{j=1}^N \big(\frac{w_{j}^{(N)}}{m_N}-\frac{\theta}{N}   \big)^2 } \big) \nonumber\\
&=& o_{P_w}(1) I_{4}(N),
\end{eqnarray}
on account of (\ref{(6.9)}) as $N,m_N \to \infty$ so that $m_N=o(N^2)$.

\par
As to the denominator  of the term $I_{4}(N) $,  we have

\begin{eqnarray*}
\sum_{j=1}^N \big(\frac{w_{j}^{(N)}}{m_N}-\frac{\theta}{N}   \big)^2 &=& \sum_{j=1}^N \Big(\big(\frac{w_{j}^{(N)}}{m_N}-\frac{1}{N}\big)+   \big(\frac{ 1}{N}-\frac{\theta}{N}\big )   \Big)^2\\
&=& \sum_{j=1}^N \big(\frac{w_{j}^{(N)}}{m_N}-\frac{1}{N}   \big)^2 + (1-\theta)^{2}/N \\
&=& \sum_{j=1}^N \big(\frac{w_{j}^{(N)}}{m_N}-\frac{1}{N}   \big)^2 \Big( 1+ \frac{(1-\theta)^2}{N(\sum_{j=1}^N \big(\frac{w_{j}^{(N)}}{m_N}-\frac{1}{N}   \big)^2-1/m_N\big)+N/m_N  } \Big) \\
&=& \sum_{j=1}^N \big(\frac{w_{j}^{(N)}}{m_N}-\frac{1}{N}   \big)^2 \Big(1+\frac{(1-\theta)^2}{N(o_{P_w}(1)+1/m_N)}  \Big),
\end{eqnarray*}
on account of Lemma \ref{Lemma 6.2}, as $N,m_N \to \infty$ so that $m_N=o(N^2)$.

\par
Consequently, asymptotically accordingly, for $I_{4}(N)$ as in (\ref{(6.14)}), we have

\begin{equation}\label{(6.15)}
I_{N}(4)=\frac{1}{1+(1/N) (\frac{(1-\theta)^2}{o_{P_w}(1)+1/m_N} )},
\end{equation}
and thus, via (\ref{(6.14)}) and (\ref{(6.15)}), we conclude that $\mathcal{H}_N(\theta)=o_{P_w}(1)$, as claimed by Lemma \ref{Lemma 6.1}. $\square$

\subsection*{Proof of Lemma \ref{Lemma 6.2}}
To prove this lemma we  note that  $E_{w}\big(\sum_{i=1}^{N} (\frac{w^{(N)}_{i}}{m_N}-\frac{1}{N})^2 \big)=\frac{(1-\frac{1}{N})}{m_N}$,  and apply   Chebyshev's inequality,  with arbitrary positive  $\varepsilon$. Accordingly, we have
\begin{eqnarray}
P\Big( \big| \sum^N_{i=1}\left( \frac{w_i^{(N)}}{m_N} - \frac{1}{N} \right)^2 - \frac{1}{m_N} \big| > \varepsilon\Big)&\leq&
\frac{m^{2}_n}{\varepsilon^{2} (1-\frac{1}{N})^2  } E_{w}\big( \sum_{i=1}^{N} (\frac{w^{(N)}_{i}}{m_n}-\frac{1}{N})^2 - \frac{(1-\frac{1}{N})}{m_n}   \big)^2 \nonumber\\
&=& \frac{m^{2}_n}{\varepsilon^{2} (1-\frac{1}{N})^2  }\ E_{w} \Big\{ \Big(\sum_{i=1}^N (\frac{w^{(N)}_i}{m_N}-\frac{1}{N})^2\Big)^2 - \frac{(1-\frac{1}{N})^2}{m^{2}_N} \Big\}^2  \nonumber\\
&=& \frac{m^{2}_N}{\varepsilon
^{2} (1-\frac{1}{N})^2  } \Big\{ N E_{w} \big( \frac{w^{(N)}_1}{m_n}-\frac{1}{N} \big)^4 \nonumber\\
&+& N(N-1) E_{w} \big[\big( \frac{w^{(N)}_1}{m_N}-\frac{1}{N} \big)^2 \big( \frac{w^{(N)}_2}{m_N}-\frac{1}{N} \big)^2\big]
- \frac{(1-\frac{1}{N})^2}{m^{2}_N}   \Big\}. \nonumber\\
&& \label{(6.16)}
\end{eqnarray}

\par
Considering that  the random weights   $w_{i}^{(N)}$ are multinomially distributed,  after some algebra one obtains the following upper bound for the right hand side of (\ref{(6.16)}):
\begin{eqnarray}
&&\frac{m^{2}_N}{\varepsilon^{2} (1-\frac{1}{N})^2  } \Big\{ \frac{1-\frac{1}{N}}{N^3 m^{3}_N } + \frac{(1-\frac{1}{N})^4}{m^{3}_N}  + \frac{(m_N -1)(1-\frac{1}{N})^2}{N m^{3}_N} +  \frac{4(N-1)}{N^3 m_N}       +\frac{1}{m^{2}_N}\nonumber\\
&&~~~~~~~~~~~~~~~~~ - \frac{1}{N m^{2}_N} + \frac{N-1}{N^{3} m^{3}_{N}} + \frac{4(N-1)}{N^2 m^{3}_{N}} - \frac{(1-\frac{1}{N})^2}{m^{2}_N}        \Big\}.   \nonumber
\end{eqnarray}
The preceding term vanishes as $N,m_N \to \infty$ so that $m_N=o(N^2)$. This observation concludes the proof of Lemma \ref{Lemma 6.2}. $\square$

%Since, as $m_N,N \to \infty$ such that $m_N=o(N)$, $\beta_{m_N,N}^{(1)}$   converge weakly to $B(.)$ on $(D,\calD, \Vert~\Vert)$ then so is  $\beta_{m_N,N}^{(2)}$. The latter, in turn, in view of (\ref{(6.6)}) implies that $\beta_{m_N,N}^{(4)}$ also converges weakly on on $(D,\calD, \Vert~\Vert)$. This completes the proof of Theorem \ref{THEOREM}. $\square$

\newpage
\section{Appendix: Confidence bands for  theoretical distributions   in combination with    virtual resampling from  large enough, or moderately small,     samples}\label{Appendix}
When all $N$ observables of large enough, or moderately small, samples are available to be processed, then
(\ref{(2.10)}) yields  the  asymptotically exact $(1-\alpha)$ size classical Kolmogorov confidence band for a continuous distribution function $F(x)$. Namely, as $N \to \infty$, with  probability $(1-\alpha)$, $\alpha \in (0,1)$, we have

\begin{equation}\label{(7.1)}
\big\{  \max\big( 0,F_N (x) - c_{\alpha}/\sqrt{N} \big) \leq F(x) \leq \min\big(1, F_{N}(x) +c_{\alpha}/\sqrt{N} \big), \ \forall x \in \IR   \Big\}
\to 1-\alpha,
\end{equation}
where $c_{\alpha}$ is as in (\ref{(3.4)}).

\par
Moreover, as noted already in Remarks \ref{Remark 2.2-1} and \ref{Remark 2.3} respectively, with $\big\{ \theta \in \IR |~ \theta\neq 1  \big\}$, in (\ref{(2.10-1)}) and, asymptotically equivalently, in (\ref{(2.16-1)}), the linear combination of the respective empirical  process as in (\ref{(1.1)}) and (\ref{(1.5)}), $\{ F_{m_N,N} (x) - \theta F_{N} (x), ~ x \in \IR \}$,  estimates $(1-\theta) F(x)$ uniformly in $x \in \IR$ (cf. (\ref{(1.11)}) and (\ref{(1.12)}) as well). In particular, via the conclusion of (\ref{(2.16-1)}), as $N,m_N \to \infty$ so that $m_N=o(N^2)$, with $\{\theta \in \IR | ~ \theta \neq 1 \}$, we arrive at having

\begin{equation}\label{(4.2)}
\big\{ L^{(3)}_{m_N,N}(x,\theta) \leq F(x) \leq U^{(3)}_{M_N,N} (x,\theta), ~ \forall x \in \IR  \big\}
\end{equation}
as an asymptotically exact $(1-\alpha)$ size confidence set for $F(.)$, both in terms of $P_{X|w}$ and $P_{X,w}$, where

\begin{equation}\label{(4.3)}
L^{(3)}_{m_N,N}(x,\theta) := \left\{
                               \begin{array}{ll}
\frac{1}{\theta-1} \Big(  \theta F_{N}(x)- F_{m_N,N}(x) \Big) - c_{\alpha} \sqrt{\frac{1}{(1-\theta)^2 m_N} + \frac{1}{N}  }, & \hbox{$\theta-1 >0$.}\\
                                 \frac{1}{1-\theta} \Big(  F_{m_N,N}(x) - \theta F_{N}(x) \Big) - c_{\alpha} \sqrt{\frac{1}{(1-\theta)^2 m_N} + \frac{1}{N}  }, & \hbox{$1-\theta >0$.}
                               \end{array}
                             \right.
\end{equation}

\begin{equation}\label{(4.4)}
U^{(3)}_{m_N,N}(x,\theta) := \left\{
                               \begin{array}{ll}
\frac{1}{1-\theta} \Big(  F_{m_N,N}(x) - \theta F_{N}(x) \Big) + c_{\alpha} \sqrt{\frac{1}{(1-\theta)^2 m_N} + \frac{1}{N}  }, & \hbox{$1-\theta >0$.}\\
                                 \frac{1}{\theta-1} \Big(   \theta F_{N}(x) - F_{m_N,N}(x) \Big)
+ c_{\alpha} \sqrt{\frac{1}{(1-\theta)^2 m_N} + \frac{1}{N}  }, & \hbox{$\theta-1 >0$.}
                               \end{array}
                             \right.
\end{equation}
and, given $\alpha \in (0,1)$, $c_\alpha$ is again as in (\ref{(3.4)}).

\par
Since $0 \leq F(x) \leq 1$, $\forall x \in \IR$, as a consequence of (\ref{(4.2)}) it follows that

\begin{equation}\label{(4.5)}
\big\{ \max \big( 0, L^{(3)}_{m_N,N}(x,\theta)  \big) \leq F(x) \leq \min\big(1,\leq U^{(3)}_{M_N,N} (x,\theta) \big), ~ \forall x \in \IR  \big\}
\end{equation}
is an asymptotically correct $(1-\alpha)$ size confidence band for $F(.)$, both in terms of $P_{X|w}$ and $P_{X,w}$, when $(1-\theta)>0$,  or $(\theta-1)>0$,  with $\{\theta \in \IR| ~ \theta\neq 1 \}$.

\par
The performance of the family of confidence bands for $F(.)$ as in (\ref{(4.5)}), indexed by $\{\theta \in \IR| ~ \theta \neq 1 \}$, is to be compared to that of the classical one as in (\ref{(4.1)}).

\begin{remark}\label{new (4.1)}
When $\theta=0$, the confidence band for $F(.)$ in (\ref{(4.5)}) coincides with that of (\ref{(3.9)}) with $L^{(2)}_{m_N,N} (.)$ and $U^{(2)}_{m_N,N}  (.)$ as in (\ref{(3.7)}) and (\ref{(3.8)}),  respectively. For $\theta=2$, (\ref{(4.3)}) and (\ref{(4.4)}) yield the bounds

\begin{equation}\label{new (4.6)}
2F_N(x) - F_{m_N,N}(x) \pm c_{\alpha} \sqrt{\frac{1}{m_N}+\frac{1}{N}}
\end{equation}
for an asymptotically exact $(1-\alpha)$  size confidence set for $F(.)$ that is to be used in (\ref{(4.5)}) for $F(.)$, whose width in this case coincides with that of (\ref{(3.9)}), as mentioned right above, when $\theta=0$. Furthermore, indexed by $\{\theta \in \IR| 0 < (1-\theta)^2< 1 \}$, i.e.,  when $\theta \in (0,2)$, the confidence bands for $F(.)$ in (\ref{(4.5)}) are wider than the ones provided by it, when taking $\theta=0$, or $\theta=2$, as above. On the other hand, the confidence bands for $F(.)$ in (\ref{(4.5)}) when indexed by $\{\theta \in \IR| (1-\theta)^2 > 1 \}$, i.e., when $\theta<0$ or $\theta>2$, are narrower than the just mentioned latter ones that are obtained via taking $\theta=0$, $\theta=2$. Thus, gaining better probability coverage via wider bands \textit{versus} wanting narrower bands is completely governed by the windows provided by $\{\theta \in \IR| \theta \in (0,2) \}$ \textit{versus} $\{\theta \in \IR| \theta <0 \}$ or $\{\theta \in \IR| \theta > 2 \}$.
\end{remark}
As illustrative examples of wider bands, with $\theta=1/2$, (\ref{(4.3)}) and (\ref{(4.4)}) yield

\begin{equation}\label{new (4.7)}
\big(2F_{m_{N},N} (x)-F_{N}(x) \big) \pm c_{\alpha} \sqrt{\frac{4}{m_N}+ \frac{1}{N}},
\end{equation}
and with $\theta=3/2$, (\ref{(4.3)}) and (\ref{(4.4)}) yield

\begin{equation}\label{new (4.8)}
\big( 3F_{N}(x)- 2 F_{m_N,N}(x) \big) \pm c_{\alpha} \sqrt{\frac{4}{m_N}+ \frac{1}{N}},
\end{equation}
as respective lower upper bounds to be used in (\ref{(4.5)}) in these two cases   for constructing the thus resulting respective confidence bands for $F(.)$.

\par
As illustrative examples of narrower bands, we mention the case of $\theta=-1$, resulting in having (\ref{(4.5)}) with respective lower and upper confidence bounds for $F(.)$
\begin{equation}\label{(4.6)}
\frac{1}{2} \big( F_{m_N,N}(x)+F_{N}(x) \big) \pm c_{\alpha} \sqrt{\frac{1}{4 m_N} + \frac{1}{N}},
\end{equation}
and the case of $\theta=3$, yielding

\begin{equation}\label{(4.7)}
\frac{1}{2} \big(3F_{N}(x) -F_{m_N,N}(x) \big) \pm c_{\alpha} \sqrt{\frac{1}{4 m_N} + \frac{1}{N}}
\end{equation}
as respective lower and upper confidence bounds for for $F(.)$   in (\ref{(4.5)}).

\begin{remark}\label{Remark 4.1}
On taking $m_N=N$ in (\ref{(4.5)}), with $\theta=0$ we obtain (\ref{(3.9)}) with $m_N=N$, and with $\theta\neq 1$ in general,  we arrive at (\ref{(4.3)}) and (\ref{(4.4)}) that are  to be used in (\ref{(4.5)}) with $m_N=N$. All these bands for $F(.)$ are to be compared  to that of the classical one as in (\ref{(4.1)}). In particular, when $\theta=0$, $F_{N,N}(.)$ estimates $F(.)$ alone, as compared to having $2F_N(.)-F_{N,N}(.)$ estimating $F(.)$ as in $(\ref{new (4.6)})$ when $\theta=2$, with the same band width, that is a bit wider than that of the classical one in (\ref{(4.1)}). The illustrative examples of even wider bands for  $F(.)$ respectively provided by the bounds as in (\ref{new (4.7)}) and (\ref{new (4.8)}) are also to be compared to each other  when $m_N=N$, as well as to that of classical case as in (\ref{(4.1)}). Similarly, the illustrative examples of having ``two-sample'' narrower bands for $F(.)$ respectively provided by (\ref{(4.7)}) and (\ref{(4.8)}) are to be compared to each other when $m_N=N$, as well as to that of the  classical one as in (\ref{(4.1)}).
\end{remark}

\begin{remark}\label{new Reamark 4.3}
On letting $m_N=N$ in (\ref{(4.3)}) and (\ref{(4.4)}) the widths of the thus obtained bands for $F(.)$ in (\ref{(4.5)}) is determined via $\pm \sqrt{\frac{1}{(1-\theta)^2 N}+ \frac{1}{N}}$, indexed by
$\{\theta \in \IR| \theta\neq 1 \}$. Thus, \textit{to begin with}, given $N\geq 1$, we may take virtual subsamples of size $m_N=(1-\theta)^2 N$, indexed by $\{\theta\in \IR | \theta\neq1 \}$, and choose desirable values for $\theta$ via $\{\theta \in \IR| ~ 0 < (1-\theta)^2 <1 \}$ in case of wanting wider confidence bands, respectively via $\{\theta \in \IR| ~ (1-\theta)^2>1 \}$ in case of wanting narrower bands, as indicated in Remark \ref{new (4.1)} and illustrated by examples thereafter.
\end{remark}

\begin{remark} \label{Remark 4.2}
The family of functionals $\sup_{x \in \IR} \big| \tilde{\beta}^{(3)}_{m_N,N}(x, \theta) \big|$ as in (\ref{(2.16-1)}), indexed by $\{\theta \in \IR| ~ \theta \neq 1 \}$, can also be used for goodness of fit tests  for $F$ against general alternatives in our present context, i.e., when all $N$ observables of large enough, or moderately small, samples  are available to be processed. Namely, for testing
$H_0:~ F=F_0$, where $F_0$ is a given continuous distribution function, we let $F=F_0$ in (\ref{(2.16-1)}), and reject $H_0$ in favor of the alternative hypothesis $H_1:~ F \neq F_0 $, for large values of the thus obtained statistic with any desirable value of $\{\theta \in \IR| ~ \theta \neq 1 \}$ at significance level $\alpha \in (0,1)$ as $N,m_N \to \infty$ so that $m_N=o(N^2)$. Thus, in view of (\ref{(2.16-1)}), this test  with the rejection region

\begin{eqnarray}\label{(4.8)}
%&&  \sqrt{\frac{N m_N (1-\theta)^2}{N+m_N (1-\theta)^2}} \sup_{x \in \IR} \Big| \frac{F_{m_N,N}(x) - \theta F_N (x)}{(1-\theta)} -F_{0}(x)      \Big|    \geq  c_{\alpha}\nonumber \\
&&\sqrt{\frac{N m_N}{N+m_N (1-\theta)^2}} \sup_{x \in \IR} \Big|  \big(F_{m_N,N}(x)- \theta F_N(x) \big) - \big( 1-\theta \big) F_{0}(x)  \Big| \geq c_{\alpha},
\end{eqnarray}
where $c_{\alpha}$ is as in (\ref{(3.4)}), is asymptotically  of size $\alpha \in (0,1)$, both in terms of $P_{X|w}$ and $P_{X,w}$, with any desirable value of $\{\theta \in \IR| ~ \theta \neq 1 \}$. With $\theta=0$, the rejection region (\ref{(4.8)}) reduces to that of (\ref{(3.10)}) as in Remark \ref{Remark 3.1}, that can, of course,  be \textit{also}  viewed and used in our present context. For values of $\{\theta \in \IR| ~ \theta \neq 1, ~ \theta \neq 0 \}$, the test in hand with   rejection region as in  (\ref{(4.8)}) can only be   used in our present context of large enough, or moderately small, samples, for then we are to compute $F_{m_N,N}(x)-\theta F_{N}(x)$ with some desirable value of $\theta $, like in (\ref{(4.6)})  or (\ref{(4.7)}), for example.
\end{remark}

\begin{remark}\label{Remark 4.3}%% It is Remark 4.5 on the pdf
Along the lines of Remark \ref{Remark 3.2}, on taking $h(.)$ to be the Cram\'{e}r-von Mises-Smirnov functional, in view of the conclusions (\ref{(2.12-1)}) and (\ref{(2.14-1)}), as $N,m_N \to \infty$ so that $m_N=o(N^2)$, we have, both in terms of $P_{X|w}$ and $P_{X,w}$,

\begin{eqnarray}\label{(4.9)}
&& \frac{N m_N  }{N+m_N (1-\theta)^2}  \int^{+\infty}_{-\infty} \Big(  F_{m_N,N}(x) - \theta F_{N}(x)  - (1-\theta) F(x)   \Big)^2 dF(x) \nonumber\\
&=& \int^{1}_{0} \Big(  \tilde{\beta}_{m_N,N}^{(3)} \big( F^{-1}(t), \theta \big)  \Big)^2 dt  \dto \int_{0}^{1} B^2(t) dt.
\end{eqnarray}
Consequently, as $N,m_N \to \infty$ so that $m_N=o(N^2)$, an asymptotic size $\alpha \in (0,1)$ Cram\'{e}r-von Mises-Smirnov test for $H_0$ versus $H_1$ as in Remark \ref{Remark 4.2} has the rejection region

\begin{equation}\label{(4.10)}
\omega_{N,m_N}^2 (\theta) := \frac{N m_N  }{N+m_N (1-\theta)^2} \int^{+\infty}_{-\infty}   \Big( F_{m_N,N} (x) -\theta F_{N} (x)- (1-\theta) F_{0}(x)  \Big)^2 dF_{0}(x) \geq \nu_{\alpha},
\end{equation}
both in terms of $P_{X|w}$ and $P_{X,w}$ with any desirable value of $\{\theta \in \IR| ~ \theta  \neq 1 \}$, where $\nu_{\alpha}$ is as in (\ref{(3.13)}). With $\theta=0$, the rejection region (\ref{(4.10)}) reduces to that of (\ref{(3.12)}), that can also be considered and used in our present context. For values of $\{\theta\in \IR|~ \theta \neq 1, ~ \theta \neq 0 \}$, just like that of (\ref{(4.8)}), the test in hand with rejection region as in (\ref{(4.10)}) can only be used in our present context of large enough, or moderately small, samples.
\end{remark}

\par
As an illustration, let $m_N=N=100$. Then,

\begin{equation}\nonumber
m_{100}= \sum_{i=1}^{100} w_{i}^{(100)}=100,
\end{equation}
where the multinomially distributed random weights $\Big(w_{1}^{(100)} , \ldots,w_{100}^{(100)} \Big)$ are generated independently from  the data $\{X_1 , \ldots, X_{100} \}$ with respective probabilities $1/100$, i.e.,

\begin{equation}\nonumber
\Big(w_{1}^{(100)} , \ldots,w_{100}^{(100)} \Big)\deq Multinomial\Big(100,1/100, \ldots,1/100\Big).
\end{equation}
These   multinomial weights, in turn, are used to  construct the asymptotic $(1-\alpha)$ size confidence bands for $F(.)$ as in Remark \ref{Remark 4.1}, and for   $F_{N}(.)$ as in (\ref{(3.5)}).

\section*{Acknowledgments} I most sincerely wish to thank my colleague Masoud M. Nasari  for his interest in, and attention to, the progress of my  preparing the presentation of the results of this paper for publication in a number of years by now. Our discussions in this regard had much helped me in the process of concluding it in its present form.


\begin{thebibliography}{99}

\bibitem{CsMANA} M. Cs\"{o}rg\H{o}, Yu. V. Martsynyuk, M. M.  Nasari,   Another look at bootstrapping the Student $t$-statistic,  Mathematical Methods of Statistics {23} (4) (2014) 256-278.

\bibitem{CSNA} M. Cs\"{o}rg\H{o}, M. M.  Nasari,    Inference from Small and big Data sets with error rates,   	    Electronic Journal of Statistics,  {9} (2015) 535-566.

\bibitem{Csorgo and Rosalsky} S. Cs\"{o}rg\H{o}, A. Rosalsky,     A survey of limit laws for bootstrapped sums.  International Journal of Mathematics and Mathematical Sciences,  45 (2003)  2835-2861.


\bibitem{Donsker} M. Donsker,  Justification and Extension of Doob's Heuristic Approach to the Kolmogorov-Smirnov Theorems, Annals of Mathematical Statistics   {23} (1952) 277-283.

\bibitem{Doob}  J. L. Doob,  Heuristic Approach to the Kolmogorov-Smirnov Theorems,  Annals of Mathematical Statistics,  {20} (1946) 393-403.

\bibitem{HartSilk} H. O. Hartley,  R. L.  Sielken Jr.,   A ``Super-Population Viewpoint'' for Finite Population Sampling,   Biometrics, {31}  (1975) 411-422.


\bibitem{Koul02} H. L. Koul,    Weighted Empirical Processes in Dynamic Nonlinear Models, (Second Edition),  Lecture Notes in Statistics, {116} (2002) Springer-Verlag New York.

\bibitem{Koul70} H. L.  Koul,  Some convergence theorems for ranks  and weighted empirical cumulatives.   Ann. Math. Statist., {41} (1970) 1768-1773.

\bibitem{KoulStaudte} H. L. Koul,  R.  Staudte,   Weak convergence of weighted empirical cumulatives based on ranks.    Ann. Math. Statist., {43} (1972)  832-841.

\bibitem{Shorack} G. R. Shorack,    The weighted empirical process of row independent random variables.   Statist. Neerlandica, {33} (1979) 169-189.

 \bibitem{ShWel} G. R. Shorack, J. A.  Wellner,  Empirical Processes with Applications to Statistics, (1986) Wiley.


\bibitem{van der Vaart} A. W. van der Vaart, J. A. Wellner, Weak Convergence and Empirical Processes, (1996) Springer, New York.
\end{thebibliography}
\end{document}